\documentclass{aa}  

\usepackage[toc,page]{appendix}
\usepackage{graphicx}
\usepackage{savesym}
\usepackage{amsmath}
\savesymbol{iint}
\usepackage{txfonts}
\restoresymbol{TXF}{iint}
\usepackage{capt-of}
\usepackage{booktabs}
\usepackage[caption=false]{subfig}
\usepackage{afterpage}
\usepackage{float}
\usepackage[colorlinks=true,urlcolor=blue,citecolor=blue,linkcolor=blue]{hyperref}
\usepackage{xcolor}
\usepackage{orcidlink}
\usepackage{multirow}
\usepackage{stfloats}

\newcommand{\kms}{\mbox{km~s$^{-1}$}}
\def\Halpha{\mbox{H$\alpha$}}
\def\Hbeta{\mbox{H$\beta$}}
\def\CaIR{\mbox{\ion{Ca}{ii}\,8542~\AA}}
\def\CaH{\mbox{\ion{Ca}{ii}\,H}}
\def\CaK{\mbox{\ion{Ca}{ii}\,K}}
\def\CaHK{\mbox{\ion{Ca}{ii}\,H\,\&\,K}}
\newcommand{\Fe}[1]{\mbox{\ion{Fe}\,\sc{#1}~\AA}}
\newcommand{\HeD}{\mbox{\ion{He}{i}\,D3}}
\usepackage[switch]{lineno} % already forced by aa.cls
\renewcommand\linenumbers{}
\renewcommand\switchlinenumbers{}
\renewcommand\rightlinenumbers{}
\renewcommand\leftlinenumbers{}

\begin{document} 

%%%%%%%%%%%%%%%%%%

\title{Synthesizing Sun-as-a-star flare spectra from high-resolution solar observations}

   \author{M.~De Wilde\inst{\ref{kuleuven}}\orcidlink{0009-0001-8236-884X}
          \and          
          A.~G.~M.~Pietrow\inst{\ref{aip},\ref{kuleuven}}\thanks{For further information, data requests, or correspondence, please contact A.~G.~M.~Pietrow at \email{apietrow@aip.de}.}\orcidlink{0000-0002-0484-7634}     
          \and
          M.~K.~Druett\inst{\ref{kuleuven}}\orcidlink{0000-0001-9845-266X}
          \and
          A.~Pastor Yabar\inst{\ref{ispstockholm}}\orcidlink{0000-0003-2359-9039} 
          \and
          J.~Koza\inst{\ref{aisas}}\orcidlink{0000-0002-7444-7046}
          \and
          I.~Kontogiannis\inst{\ref{ethz},\ref{irsol}}\orcidlink{0000-0002-3694-4527}
          \and
          \\
          O.~Andriienko\inst{\ref{ispstockholm}}
          \and
          A.~Berlicki\inst{\ref{wroc},\ref{ond}}\orcidlink{0000-0002-6505-4478}
          \and
          A.~R.~Brunvoll\inst{\ref{oslo1},\ref{oslo2}}\orcidlink{0009-0004-6326-5652}
          \and
          J.~de la Cruz Rodr\'{\i}guez\inst{\ref{ispstockholm}}\orcidlink{0000-0002-4640-5658}
          \and
          J.~T.~Faber\inst{\ref{oslo1},\ref{oslo2}}\orcidlink{0009-0006-6837-1504}
          \and
          R.~Joshi\inst{\ref{fairfax},\ref{nasa},\ref{oslo2}}\orcidlink{0000-0003-0020-5754}
          \and
          D.~Kuridze\inst{\ref{col},\ref{geor}}\orcidlink{0000-0003-2760-2311}
          \and
          D.~N\'{o}brega-Siverio\inst{\ref{iac1},\ref{iac2},\ref{oslo1},\ref{oslo2}}\orcidlink{0000-0002-7788-6482}
          \and
          L.~H.~M.~Rouppe van der Voort\inst{\ref{oslo1},\ref{oslo2}}\orcidlink{0000-0003-2088-028X}
          \and
          J.~Ryb\'{a}k\inst{\ref{aisas}}\orcidlink{0000-0003-3128-8396}
          \and
          E.~Scullion\inst{\ref{nu}}\orcidlink{0000-0001-9590-6427}
          \and
          A.~M.~Silva\inst{\ref{up},\ref{up2}}\orcidlink{0000-0003-4920-738X}          
          \and
          Z.~Vashalomidze\inst{\ref{aisas}}\orcidlink{0009-0002-6790-8796} 
          \and
          A.~Vicente Ar\'{e}valo\inst{\ref{kis}}\orcidlink{0000-0003-3896-836X}
          \and
          A.~Wi\'{s}niewska\inst{\ref{aisas}}\orcidlink{0000-0002-7805-0732}
          \and
          R.~Yadav\inst{\ref{col}}\orcidlink{0000-0003-4065-0078}
          \and
          T.~V.~Zaqarashvili\inst{\ref{ug},\ref{isu},\ref{geor}}\orcidlink{0000-0001-5015-5762}
          \and
          J.~Zbinden\inst{\ref{ub}}\orcidlink{000000-0003-2780-7851}
          \and
          E.~S.~{\O}yre\inst{\ref{oslo1},\ref{oslo2}}\orcidlink{0009-0005-6122-1987}
          }

\institute{
Centre for Mathematical Plasma Astrophysics, KU Leuven, Celestijnenlaan 200B, B-3001 Leuven, Belgium\label{kuleuven}
\and
Leibniz-Institut f\"{u}r Astrophysik Potsdam (AIP), An der Sternwarte 16, 14482 Potsdam, Germany\label{aip}
\and
Inst. for Solar Physics, Dept. of Astronomy, Stockholm University, Albanova University Centre, SE-106 91 Stockholm, \mbox{Sweden}\label{ispstockholm}
\and
Astronomical Institute, Slovak Academy of Sciences, Tatransk\'{a} Lomnica, 059 60, Slovakia\label{aisas}
\and
ETH Z\"{u}rich, Institute for Particle Physics and Astrophysics, Wolfgang-Pauli-Strasse 27, 8093, Z\"{u}rich, Switzerland\label{ethz}
\and
Istituto Ricerche Solari Aldo e Cele Dacc\'{o} (IRSOL) Locarno, Switzerland\label{irsol}
\and
Centre of Scientific Excellence – Solar and Stellar Activity, University of Wroc\l{}aw, 50137 Wroc\l{}aw Poland\label{wroc}
\and
Astronomical Institute of the Czech Academy of Sciences, 251 65 Ond\v{r}ejov, Czech Republic\label{ond}
\and
Institute of Theoretical Astrophysics, University of Oslo, P.O. Box 1029 Blindern, N-0315, Oslo, Norway\label{oslo1}
\and
Rosseland Centre for Solar Physics, University of Oslo, P.O. Box 1029 Blindern, N-0315 Oslo, Norway\label{oslo2}
\and
Department of Physics and Astronomy, George Mason University, Fairfax, VA 22030, USA\label{fairfax}
\and
Heliophysics Science Division, NASA Goddard Space Flight Center, Greenbelt, MD 20771, USA\label{nasa}
\and
National Solar Observatory, 3665 Discovery Drive, Boulder, CO 80303, USA\label{col}
\and
Evgeni Kharadze Georgian National Astrophysical Observatory, Mount Kanobili, 0301 Abastumani, Georgia\label{geor}
\and
Instituto de Astrof\'{\i}sica de Canarias, E-38205 La Laguna, Tenerife, Spain\label{iac1}
\and
Universidad de La Laguna, Dept. Astrof\'{\i}sica, E-38206 La Laguna, Tenerife, Spain\label{iac2}
\and
Northumbria University, NE1 8ST Newcastle upon Tyne, UK\label{nu}
\and
Instituto de Astrof\'{\i}sica e Ci\^{e}ncias do Espa\c{c}o, CAUP, Universidade do Porto, Rua das Estrelas, 4150-762, Porto, Portugal\label{up}
\and
Departamento de Fisica e Astronomia, Faculdade de Ciencias, Universidade do Porto, Rua do Campo Alegre, Porto, Portugal\label{up2}
\and
Institut f\"{u}r Sonnenphysik (KIS), Georges-K\"{o}hler-Allee 401 A, Freiburg i.Br., Germany\label{kis}
\and
Institute of Physics, University of Graz, Universit\"{a}tsplatz 5, 8010, Graz, Austria\label{ug}
\and
Department of Astronomy and Astrophysics at Space Research Center, Ilia State University, Tbilisi, Georgia\label{isu}
\and
Astronomical Institute of the University of Bern, Sidlerstrasse 5, 3012 Bern, Switzerland\label{ub}
}

\date{Accepted: July, 09, 2025}

\abstract
{Spatially resolved observations of the Sun and the astronomical sample size of stellar bodies are the respective key strengths of solar and stellar observations. However, the large difference in object brightness between the Sun and other stars has led to distinctly different instrumentation and methodologies between the two fields.} 
{We produced and analyzed synthetic \mbox{full-disk} spectra derived from 19 small area \mbox{field-of-view} optical observations of solar flares acquired by the Swedish 1-m Solar Telescope (SST) between 2011 and 2024. These were used to investigate what can and cannot be inferred about physical processes on the Sun from \mbox{Sun-as-a-star} observations.}
{The recently released Numerical Empirical Sun-as-a-Star Integrator (NESSI) code provides synthetic \mbox{full-disk} integrated spectral line emission based on smaller \mbox{field-of-view} input while accounting for \mbox{center-to-limb} variations and differential rotation. We used this code to generate pseudo-Sun-as-a-star spectra from the SST observations.}
{We show that limited-area solar observations can be extrapolated to represent the full disk accurately in a manner close to what is achievable with \mbox{Sun-as-a-star} telescopes. Additionally, we identify six spectral features, two of which are caused by instrumental effects. Most notably, we find a relation between the heliocentric angle of flares and the width of the excess emission left by them as well as a source of false positive \mbox{coronal mass ejections-like} signatures, and we defined an energy scaling law based on chromospheric line intensities that shows that the peak flare contrast roughly scales with the square root of the bolometric energy.}
% conclusions
{The presented method of creating \mbox{pseudo-Sun-as-a-star} observations from limited \mbox{field-of-view} solar observations allows for the accurate comparison of solar flare spectra with their stellar counterparts while allowing for the detection of signals at otherwise unachievable noise levels.}


\maketitle

\clearpage

%%%%%%%%%%%%%%%%%%%
%%%    Fig 1   %%%
%%%%%%%%%%%%%%%%%%%
\begin{figure*}[t!]
    \centering
    \includegraphics[width=\textwidth]{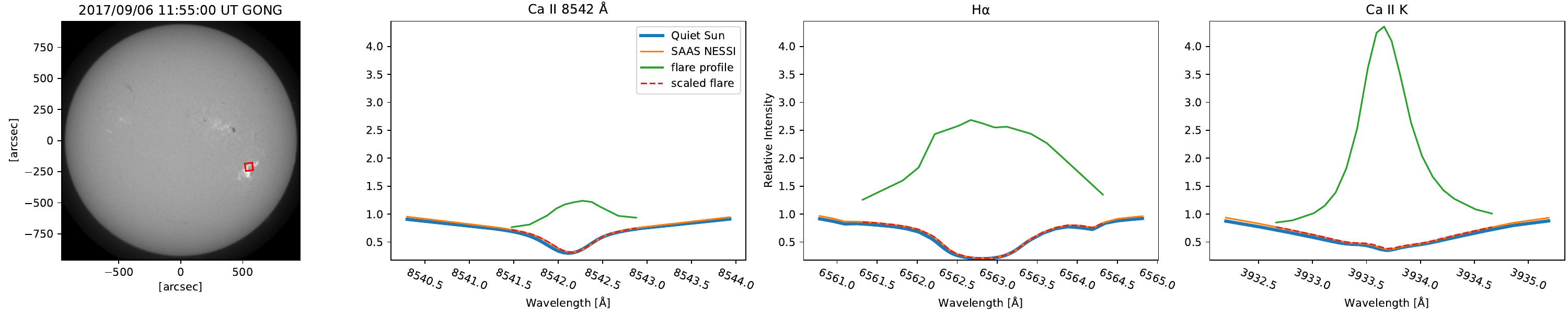}
    \caption{Comparison of localized flare profiles and disk-integrated Sun-as-a-star profiles. Left: A GONG \Halpha\ line core image showing a large solar flare (red box) within a field of view (FOV) typical of high-resolution solar observations. Right: The \CaIR, \Halpha, and \CaK\ line profiles at different locations: within the FOV in a quiet region (blue), integrated over the quiet-Sun disk for the Sun as a star (orange), and on the flare ribbon (green). The scaled flare (red dashed) is a simple ratio scaling of the FOV area to the total disk area, while the NESSI profile is the result of injecting the average flare profile (green) into a full-disk quiet-sun profile (blue). All profiles are normalized to their respective quasi-continua. The small difference between the red dashed and blue lines illustrates the stark differences between solar and stellar observations.}
    \label{fig:NESSI}
\end{figure*}
%%%%%%%%%%%%%%%%

\section{Introduction} 
\label{sec:intro} 

Solar flares and the corresponding coronal mass ejections (CMEs) are the primary drivers of the most severe disruptions to the heliosphere \citep{2022Cliver}. When aimed at the Earth, they can cause widespread damage to electronics, blackouts, and disruptions in communications  \citep[e.g.,][]{Cole2003,Miteva2023}. The energies of these flares are typically between $10^{29}$ and $10^{33}$\,erg \citep{2011Fletcher_flare_review_SSRv..159...19F}, which is still two orders of magnitude lower than the most powerful stellar flares. These extreme events are believed to reach energies as high as $10^{35}$ or $10^{36}$\,erg, depending on whether one adopts the assumption of blackbody radiation \citep{2012Maehara} or hydrogen recombination continuum \citep{2024Simoes} for energy fitting. While solar flares leave only faint \citep{woods2006,Kretzschmar2011} or at times no detectable \citep[e.g.,][]{2024pietrowHarps} signatures in the total solar irradiance, they can dramatically outshine their host stars when occurring on intrinsically dimmer M~dwarfs \citep[e.g.,][]{Zhan2019}. They are also visible as dimmings in the corona and transition region \citep[e.g., ][]{Hou2025,Veronig2025}. 

Although such extreme events have been observed on other stars for years \citep[e.g.,][]{Pietras2022,Namizaki2023,Bruno2024}, stellar flares have only recently attracted significant attention in the context of exoplanet research. In part because they are sources of noise and transit time variation \citep[e.g.,][]{Davenport2019,Howard2023} but also since they are agents of change in planetary environments, where it is believed that flares and their ejecta can permanently alter planetary atmospheres and thus their habitability \citep[e.g.,][]{Pulkkinen2007,Airapetian2020,Temmer2021, Ilin2024}. 

For this reason, it is important to understand stellar flares and what differentiate them from their solar counterparts. However, this is not straightforward, as solar flares are typically studied in high spectral and spatial resolution, while the opposite is true for stellar flares, which are primarily observed in white light, and only more recently with \mbox{high-resolution} spectrographs \citep[e.g.,][]{Namekata2021,Martinez2022,Namizaki2023}. As solar flares imprint only faint \citep{woods2006,Kretzschmar2011} or non-detectable \citep{2024pietrowHarps} \mbox{white-light} fluctuations on a \mbox{disk-integrated} scale, direct comparisons have so far only been possible within numerical models \citep[e.g.,][]{Alred2005,2024_Druett_amrvac_flare_A&A...684A.171D}.
However, this is no longer the case when looking at spectral lines where even relatively weak flares for stellar standards can leave an imprint on a disk-integrated spectrum (see Fig.~\ref{fig:NESSI}). 

\citet{Namekata2021, Namekata2022Saas}, \citet{Otsu_2022}, and \citet{Xu2022} pioneered a \mbox{pseudo-Sun-as-a-star} analysis technique for flares and CMEs by studying line-profile evolutions of spatially integrated spectra of a small region around the event. Here, a difference was seen between the imprint of the flare on the spectra and its position on the disk. Such a relation has already been known to exist between UV and \mbox{X-ray} radiation for solar flares \citep{woods2006} and has recently been suggested to also exist for spectral lines in the visible by \citet{Yu2025} through 1D flare simulations using the RADYN code \citep{Alred2005}. If confirmed in observations, such an effect could prove to be a powerful technique for determining the disk position of these events.  

However, it was not clear if these patterns found using the \mbox{pseudo-Sun-as-a-star} technique are truly related to the flare location, or whether this is a result of small \mbox{field-of-view} (FOV) integrations, which do not take into account the \mbox{center-to-limb} variations (CLV) and differential rotation of the rest of the star. Its validity was largely confirmed by the first
true \mbox{Sun-as-a-star} flare analysis \citep{2024pietrowHarps} that used the solar telescope \citep{Dumusque(2015)} connected to the High Accuracy \mbox{Radial-velocity} Planet Searcher for the Northern hemisphere \citep[\mbox{HARPS-N};][]{Cosentino2012} instrument on the Telescopio Nazionale Galileo. 

Follow-up studies with such \mbox{Sun-as-a-star} telescopes are limited by their \mbox{signal-to-noise} ratio (S/N) and are primarily sensitive to very strong flares. On the other hand, there is a large collection of flare observations taken with high spatial and spectral resolution in the solar community that have excellent S/N but a limited FOV. Such observations can be processed in a way similar to that of other \mbox{pseudo-Sun-as-a-star} studies to obtain more statistics on flare behavior in a \mbox{disk-integrated} setting. However, we propose adding an additional step to more closely approximate true \mbox{Sun-as-a-star} flare observations. In this work, we apply the Numerical Empirical \mbox{Sun-as-a-Star} Integrator \citep[NESSI;][]{2023PietrowNESSI} code to synthesize \mbox{Sun-as-a-star} emission from 20 solar flare observations using data with a limited FOV, obtained by the Swedish 1-m Solar Telescope \citep[SST;][]{Scharmer03}. This work also serves as a pilot study on how the Paranal solar Espresso Telescope \citep[POET;][]{Leite2024, Santos2025} can observe flares within its limited FOV, ranging from 1\arcsec\ to 60\arcsec, and therefore also explores what can be inferred from partially captured flares.

\section{Data and methods}
\label{sec:method} 

\begin{figure}
    \centering
    \includegraphics[width=\linewidth]{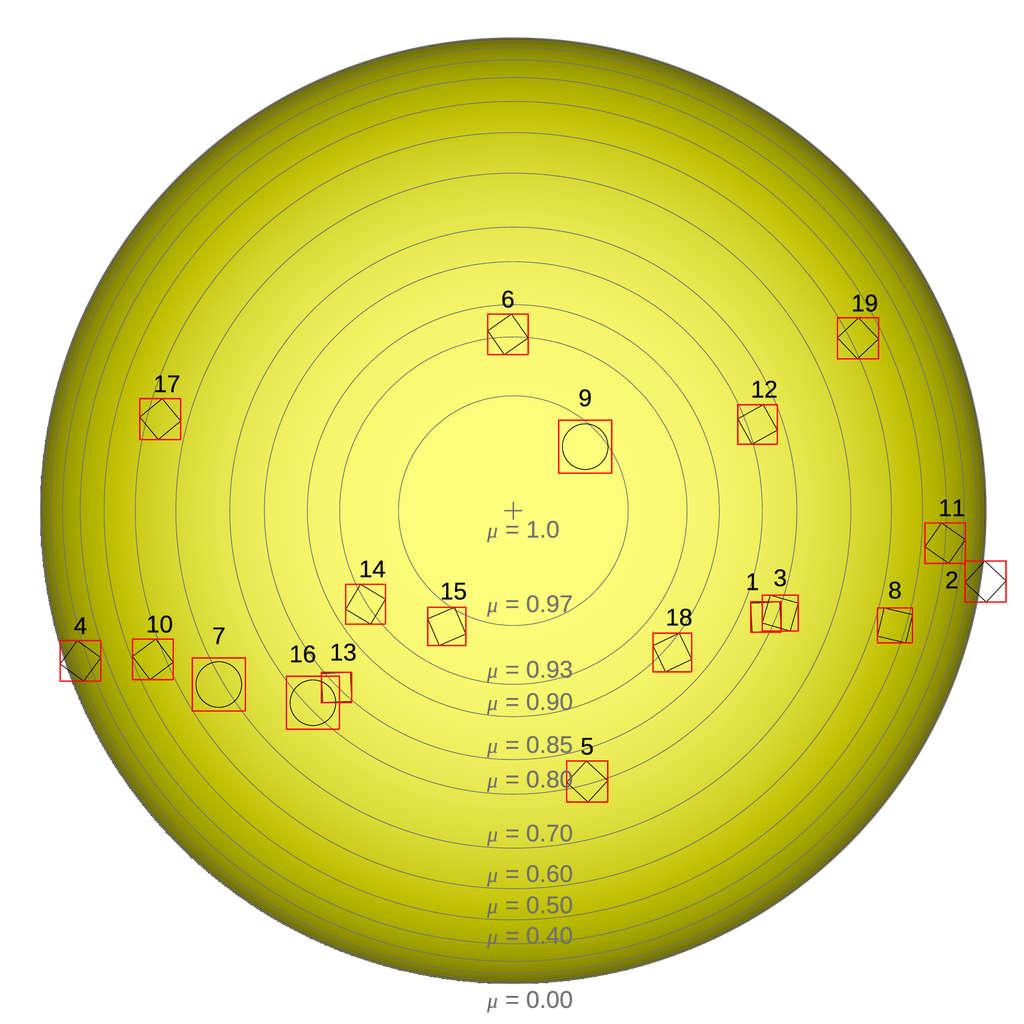}
    \caption{Flare locations on the solar disk (red boxes) labeled with event numbers from Table~\ref{tab:flare_info}. Tilted black squares and black circles (corresponding to Events~7, 9, and 16) approximately match the CRISP or CHROMIS FOV in the COCOPLOTS in Figs.~\ref{fig:X9.3}–\ref{fig:C1.2}. Gray concentric circles correspond to directional cosines $\mu$, labeled accordingly, with the disk center marked by a plus sign.}
    \label{fig:sun_with_observations}
\end{figure}

In this work, we made an effort to collect and discuss a statistically significant number of 20 solar flares observed by the SST. These flares were observed in various lines, using the CRisp Imaging SpectroPolarimeter \citep[CRISP;][]{Scharmer08} and the CHROMospheric Imaging Spectrometer \citep[CHROMIS;][]{Scharmer17}. Both instruments are based around a set of two  Fabry-P\'{e}rot interferometers, allowing for accurate wavelength tuning, and thus can "scan" through a spectral line to get a \mbox{quasi-simultaneous} spectrum at an a\-ve\-rage spectral resolution ${\cal R} \approx 130\,000$. The flare locations are shown in Fig.~\ref{fig:sun_with_observations} and the flare characteristics are summarized in Table~\ref{tab:flare_info}.

The data used in this work are a collection of observations spanning the period from 2011 to 2024, collected by various institutes for different scientific purposes. As a result, the observations use different line samplings, cadences, and pointing strategies. All sets are reduced by some version of the standard CRISPRED and SSTRED pipelines \citep{Jaime2015, Mats2021}, but there are differences in the output products. Older cubes were saved in the legacy "La Palma cube" format and contain polarimetric fringes and poor intensity calibration. These cubes were also not always rotated in such a way that solar north is up. In these cases, the rotation was performed manually based on the SST turret logs, intensity calibrations were performed, and the cubes were saved into the standard FITS format.

Furthermore, all data is reconstructed using \mbox{multi-object} \mbox{multi-frame} blind deconvolution \citep[MOMFBD;][]{mats02,vanNoort05}, which is known to introduce artifacts in frames with very bad seeing conditions \citep[see, e.g., Fig.~5.4 of][]{Pietrowthesis}. This is one reason why observations under such conditions are typically avoided. However, solar flare observations are sufficiently rare that exceptions are made in this case. The MOMFBD artifacts exacerbate the effects of poor seeing on \mbox{FOV-averaged} spectra and cannot be corrected because of the legacy origins of the dataset. This leaves recognizable signatures in the spectra, which we discuss and point out in Section~\ref{sec:results}.

Additionally, each flare is checked for jumps in tracking and bad pixels, and where possible, the data is cropped accordingly to remove these effects. With this, we aim to minimize any instrumental effects on the \mbox{Sun-as-a-star} analysis. 

The strengths of solar flares are characterized using the scale by \citet{1970BakerGOESClassification}, which uses the peak flux measured in the Geostationary Operational Environmental Satellite \citep[GOES;][]{1994GOES} $1-8$~\AA\ channel to divide flares into one of five classes (A, B, C, M, and X class) where each class represents an order of magnitude jump in flux. While \citet{Warmuth2016} have shown that this scale does not scale one-to-one with radiated bolometric flare energy, this is generally assumed. 

In this case, an X1.0 flare is defined as having $10^{31}$~erg (e.g., Fig.~2 in \citet{Shibata2013} and Fig.~5 in \citet{Maehara2015}). As flares of this type are generally rare, this indicates that the Sun's flare output is significantly below average, as \citet{Pietras2022} report an average stellar flare strength of approximately X100 (or X10 based on the assumptions of \citet{2024Simoes}) for \mbox{G-type} stars.

%%%%%%%%%%%%%%%%
%%%  Fig 2   %%%
%%%%%%%%%%%%%%%%

%%%%%%%%%%%%%%%%
%%%  Fig 3   %%%
%%%%%%%%%%%%%%%%
%%% Proof-of-Concept Figure
\begin{figure*}
    \centering
\includegraphics[width=0.75\textwidth]{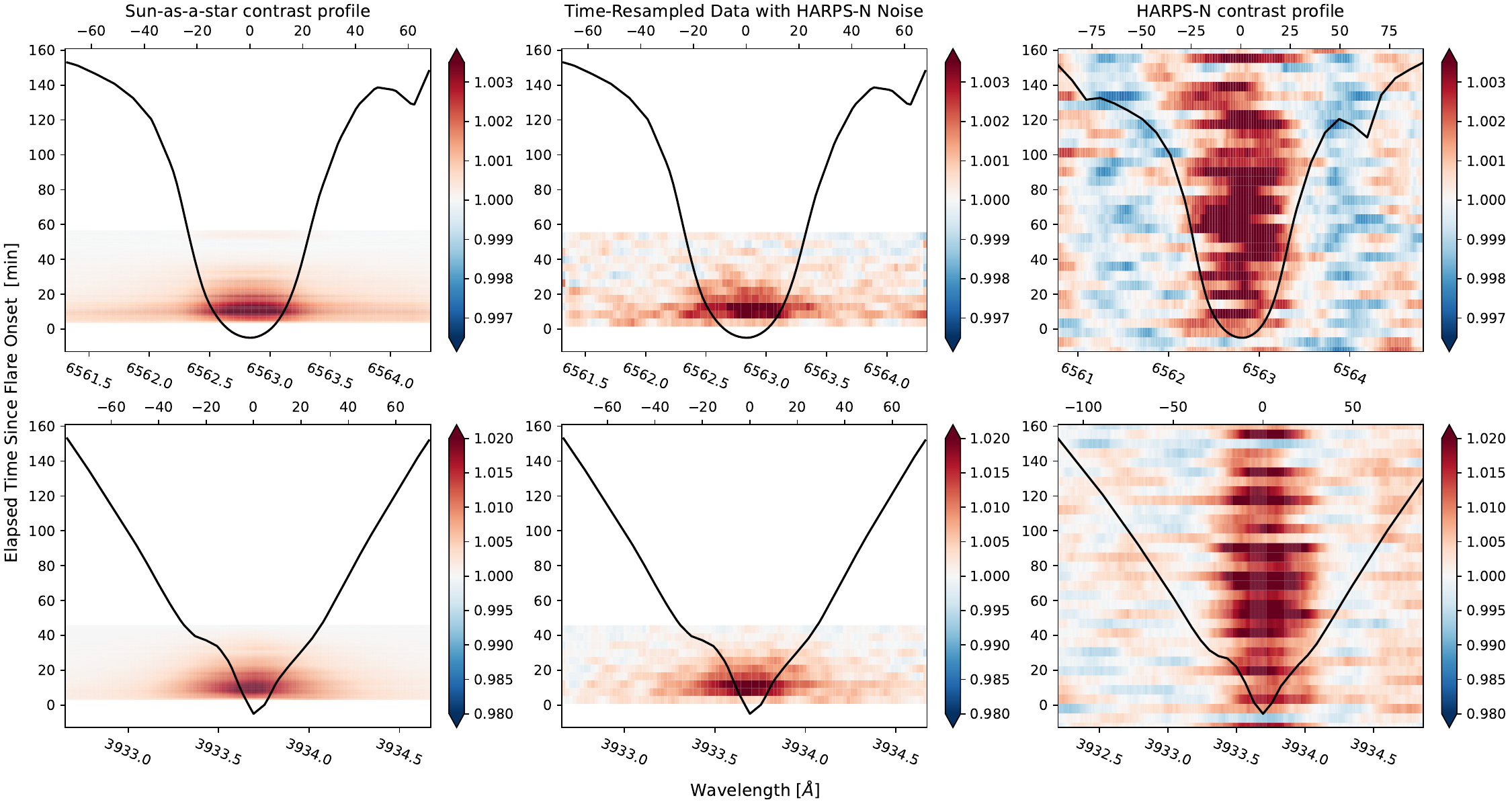}
    \caption{Construction process of the temporal variation of the \mbox{Sun-as-a-star} contrast profile, $C(\lambda, t)$ (Eq.~\ref{eq:contrast_profile}), with a resolution equivalent to \mbox{HARPS-N}, derived from \mbox{high-resolution} SST observations of the X9.3 flare, listed as Event~1 in Table~\ref{tab:flare_info}, for \Halpha\ (top row) and \CaK\ (bottom row). Each panel presents the contrast profile relative to \mbox{quiet-Sun} profiles (black curves). Color bars denote the values of $C(\lambda, t)$. Left column: \mbox{Sun-as-a-star} contrast profiles generated using the NESSI code. Middle column: Same as the left column but resampled to the \mbox{HARPS-N} temporal resolution of 5 minutes and degraded by adding normally distributed noise with a standard deviation estimated from the wings of line profiles observed by \mbox{HARPS-N}. Right column: Contrast profiles of \Halpha\ and \CaK\ reconstructed from \mbox{HARPS-N} observations of Event~1, shown also in Fig.~\ref{fig:X9.3}.}
    \label{fig:X9.3HARPSnoise}
\end{figure*}
%%%%%%%%%%%%%%%%

\subsection{NESSI}

The NESSI code \citep[][]{2023PietrowNESSI} was created to convert limited FOV solar observations into simulated \mbox{full-disk} \mbox{Sun-as-a-star} observations. This is done using accurate CLV measurements for a given wavelength range to create a solar disk on a polar grid. The line profiles on this grid are then shifted according to a differential rotation law, such as the one presented in \citet{Balthasar1986} and assumed to be constant for all lines. This grid is then integrated to create a \mbox{quiet-Sun} profile of the \mbox{Sun-as-a-star}. Flare observations can be injected by dividing their average spectra by a subfield of the NESSI \mbox{quiet-Sun} disk, matching the size and position of the SST FOV,  after which the resulting spectrum is multiplied with the \mbox{quiet-Sun} disk. This makes it as if the SST flare had occurred on the full disk. The process is illustrated in Fig.~\ref{fig:NESSI}. NESSI works in a similar way to the Spot Oscillation And Planet code \citep[SOAP;][]{Boisse2012, Dumusque2014} but relies fully on empirical data. 

This approach of creating \mbox{pseudo-Sun-as-a-star} observations is similar to the one presented by \citet{Otsu_2022}, where small areas on the disk are integrated and normalized by \mbox{quiet-Sun} areas of similar size elsewhere on the disk. However, our approach allows for the preservation of intensities, thus showing how bright and detectable a flare would be in a true \mbox{Sun-as-a-star} observation such as the one presented in \citet{2024pietrowHarps}. This method is a digital analog to the observational strategy of the POET telescope, and many of the issues discussed in this work will be relevant to future observations with this telescope. The codes used in the \mbox{follow-up} analysis, the example images, and animations are \mbox{publicly available} at \href{https://github.com/demichieldewilde/solar-flares}{this repository}\footnote{\label{foot:git}\href{https://github.com/demichieldewilde/solar-flares}{https://github.com/demichieldewilde/solar-flares}}, as well as included with this paper.

\subsection{Data normalization}

We determine the spectral contribution of each FOV by comparing the dataset to a quiet reference region contained within. For each observation, an area containing the "quietest Sun" is manually selected inside one of the frames of the time series, away from the peak of the flare.  Intensity data over this selected area are averaged to obtain the reference \mbox{quiet-Sun} spectral intensity\footnote{In the following, quantities representing spectral intensity are distinguished from specific intensities $I$ by the subscripts $_\mathrm{QR}$, $_\mathrm{FOV}$, and $_\mathrm{FD}$, which indicate area integrals over a \mbox{Quiet-Sun} Region, the Field of View, or the Full Disk, respectively.} as a function of wavelength $I_\mathrm{QR}(\lambda)$. The NESSI code is then used to generate a theoretical \mbox{quiet-Sun} profile at this disk location $I_\mathrm{QR}^\mathrm{(N)}(\lambda)$, and the two are scaled to calibrate the observational data to the NESSI disk data. The continuum wavelength is used where available, and otherwise the furthest wing wavelength points are used, as these are less sensitive to flares and other activity signals. The scaling constant $\alpha$ is introduced such that the merit function 
%%%%%%%%%%%%%%%%%%%%%%%%%%%%%%%%%%%%%%%%%%%%%%
%%%            Equation 1                  %%%
%%%%%%%%%%%%%%%%%%%%%%%%%%%%%%%%%%%%%%%%%%%%%%
\begin{equation}
   \chi^2 = \sum_j \left(I_\mathrm{QR}^\mathrm{(N)}(\lambda_j) - \alpha I_\mathrm{QR}(\lambda_j) \right)^2 
\end{equation} 
%%%%%%%%%%%%%%%%%%%%%%%%%%%%%%%%%%%%%%%%%%%%%%
is minimized. This constant $\alpha$ is then used to normalize the data. Hence, the NESSI spectrum can now be directly compared and subtracted from the observed spectra.

We took great caution choosing the quiet reference regions because generally the flares occurred in the vicinity of a sunspot or AR, which therefore is not truly a quiet Sun. 
This is an intrinsic limitation of small FOV observations. Nevertheless, using a gauge to a theoretical quiet Sun profile makes the method more robust. Further, the evolutionary timescales of spots and other activity in the FOV are much slower than those of the flares. 
Therefore, their influence in the contrast profile (see below) will largely be mitigated by dividing out the reference spectrum.
Further confounding factors caused by using small FOV observations include variations in seeing conditions, and bright (or dark) areas moving into or out of the FOV caused by jumps in pointing.

\subsection{Contrast profiles}
\label{sec:contrast}

To study the impact of a flare on an otherwise quiet Sun, we calculated the difference between the observed integrated FOV spectrum and the NESSI spectrum. Both are averaged and subtracted so that
%%%%%%%%%%%%%%%%%%%%%%%%%%%%%%%%%%%%%%%%%%%%%%
%%%              Equation 2                %%%
%%%%%%%%%%%%%%%%%%%%%%%%%%%%%%%%%%%%%%%%%%%%%%
\begin{equation}
\label{eq:delta_I}
\Delta I_\mathrm{FOV}(\lambda, t) = \frac{O^{-1}}{A_\mathrm{FOV}} \int_\mathrm{FOV} I(x,y,\lambda,t) - I^\mathrm{(N)}(x,y,\lambda)\,\mathrm{d}x\mathrm{d}y\,,
\end{equation}
%%%%%%%%%%%%%%%%%%%%%%%%%%%%%%%%%%%%%%%%%%%%%%
where $A_\mathrm{FOV}$ is the FOV area and $O$, shown in Table~\ref{tab:flare_info}, is the fractional overlap estimated mainly between the CRISP FOV and the total flare area. This factor is introduced to approximate the full flare intensity as accurately as possible in the assumption of a linear relation between coverage and intensity. For events with significantly limited spatial coverage in the CHROMIS FOV, the overlap was averaged between the CRISP FOV and the CHROMIS FOV relative to the total flare area. All quantities will be corrected for partial coverage.

This spectral difference is added onto the NESSI \mbox{full-disk} (FD) spectrum, effectively replacing the area that has been subtracted from the observed FOV,
%%%%%%%%%%%%%%%%%%%%%%%%%%%%%%%%%%%%%%%%%%%%%%
%%%              Equation 3                %%%
%%%%%%%%%%%%%%%%%%%%%%%%%%%%%%%%%%%%%%%%%%%%%%
\begin{equation}
I_\mathrm{FD}(\lambda,t) = I^\mathrm{(N)}_\mathrm{FD} (\lambda ) + \Delta I_\mathrm{FOV}(\lambda, t)\,.
\end{equation}
%%%%%%%%%%%%%%%%%%%%%%%%%%%%%%%%%%%%%%%%%%%%%%
Therefrom the contrast profile can be defined as 
%%%%%%%%%%%%%%%%%%%%%%%%%%%%%%%%%%%%%%%%%%%%%%
%%%              Equation 4                %%%
%%%%%%%%%%%%%%%%%%%%%%%%%%%%%%%%%%%%%%%%%%%%%%
\begin{equation}
C(\lambda, t) = I_\mathrm{FD}(\lambda,t) / I^\mathrm{(N)}_\mathrm{FD} (\lambda)\,. 
\label{eq:contrast_profile}
\end{equation}
%%%%%%%%%%%%%%%%%%%%%%%%%%%%%%%%%%%%%%%%%%%%%%
This method is similar to the one of \citet{Otsu_2022} but with the added benefit of preserving the \mbox{full-disk} intensities, thus allowing the measurement of realistic contrast profiles. Many of these will be far below the detection limit of current \mbox{Sun-as-a-star} instruments. 

\subsection{Noise simulations}

We can test how well the synthesized spectra fit \mbox{full-disk} observations by degrading the SST spectra to the resolution of \mbox{HARPS-N} and matching the noise profiles. 

We made a case study based on observations of the X9.3 flare, where we compare our synthetic \mbox{Sun-as-a-star} contrast profiles with those of the same flare, which was also observed with the \mbox{HARPS-N} telescope \citep{2024pietrowHarps}. Firstly, we match the 5-minute cadence by taking every 20th observation. We do not change the spectral resolution, as the ${\cal R} \approx 130\,000$ resolution of SST/CRISP is very similar to the ${\cal R} \approx 125\,000$ resolution of \mbox{HARPS-N}. The S/N of the observations are matched next, where we add noise based on the standard deviation of the \mbox{HARPS-N} continuum close to the \CaK\ and \Halpha\ lines. By matching the noise level of \mbox{HARPS-N} both photon and instrumental noise is accounted for. Finally, the resulting profile is multiplied by 2.5 to account for the fact that we capture only about 40\% of this flare in our FOV. 

The resulting contrast profiles for \Halpha\ and \CaK\ are shown in Fig.~\ref{fig:X9.3HARPSnoise} in the top and bottom row, respectively. For both lines, the first column shows contrast profiles of the undegraded SST contrast profiles. The second column shows the same profiles degraded to the \mbox{HARPS-N} cadence and S/N, and the final column shows the corresponding \mbox{HARPS-N} profiles. 

While the SST data was observed for a much shorter time, we find a strong similarity with the contrast profile shapes and intensities between minutes 10 and 20. After this time, the flare seemingly gets dimmer in the SST frame while it stays bright in the other data. This is consistent with the flaring activity inside and outside of the limited FOV of the SST observations, which captures only around 40\% of the flare, i.e., toward the middle of flare's lifetime, most of the bright ribbon feature lies outside of the $60\arcsec \times 60\arcsec$ SST FOV. 

\subsection{Differenced equivalent width}

The differenced equivalent width of disk- or \mbox{FOV-integrated} flare observations has been shown to be correlated with \mbox{large-scale} flare evolution as it captures the line broadening and Doppler effects \citep{Namizaki2023,Otsu_2022}. 

In \citet{Otsu_2022} a quiet FOV of the same size and $\mu$ value is used as a \mbox{quiet-Sun} gauge. We instead use the \mbox{NESSI-generated} FOV at the exact flare location, thus avoiding issues with differential rotation. This method works particularly well in regions close to the limb, where a strong gradient in $\mu$ is present within the FOV.

\begin{figure*}
\centering
\includegraphics[width=\textwidth]{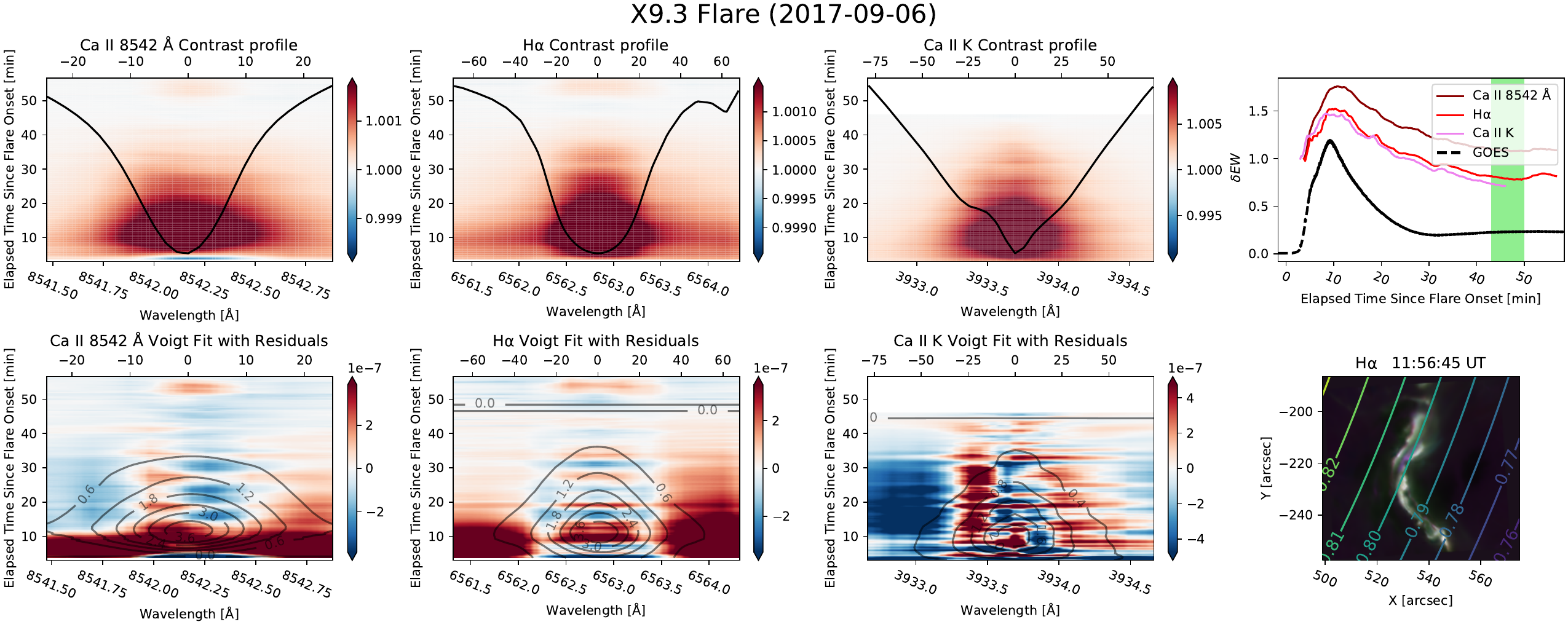}
\caption{Event~1. Top row: Temporal variations of the \mbox{Sun-as-a-star} contrast profile, $C(\lambda, t)$ (Eq.~\ref{eq:contrast_profile}), for the \CaIR, \Halpha, and \CaK\ lines, respectively. Each panel shows the contrast profile relative to \mbox{quiet-Sun} profiles (black curves) and the relative differenced equivalent widths $\delta EW$ (top right panel), integrated over $\Delta\lambda = \pm 0.5$\,\AA\ around the line centers (Eq.~\ref{eq:EW}) and normalized relative to the $|\Delta EW(t_0)|$ value of particular line at the SST Start time $t_0$ (Table~\ref{tab:flare_info}). The green bar marks the reference time span $\Delta t$, which serves as the reference for the contrast profile $C(\lambda, t)$. The GOES \mbox{X-ray} flux in the $1\text{–}8$\,\AA\ range is overplotted after ad hoc scaling. Color bars indicate the values of $C(\lambda, t)$. Bottom row: Temporal variations of the model function  (Eq.~\ref{eq:fit}) for the \CaIR, \Halpha, and \CaK\ lines, represented by black contours, plotted over the residuals $R(\lambda, t)$ of the fits (Eq.~\ref{eq:residuals}). The contours are labeled by spectral intensities normalized to the \mbox{quiet-Sun} continuum spectral intensity. The contour labeled 0 approximately corresponds to the midpoint of the time span $\Delta t$, represented by the green bar in the $\delta EW$ panel. The dashed contours indicate negative values. Color bars indicate the values of $R(\lambda, t)$. The top $x$-axes display velocities in \kms\ corresponding to Doppler shifts relative to the line centers. Bottom right panel: The cocoplot overlaid with contours of the respective directional cosines $\mu$ \citep{druett_cocoplot_2022}.}
\label{fig:X9.3}
\end{figure*}

\citet{Otsu_2022} defined their normalized spectral change for a specific time step $t_0$ as 
%%%%%%%%%%%%%%%%%%%%%%%%%%%%%%%%%%%%%%%%%%%%%%
%%%              Equation 5                %%%
%%%%%%%%%%%%%%%%%%%%%%%%%%%%%%%%%%%%%%%%%%%%%%
\begin{equation}
\Delta S(\lambda,t ) = \frac{I_\mathrm{FOV}(\lambda,t) - I_\mathrm{QR}(\lambda, t_0)}{I_\mathrm{FD}(\lambda_\mathrm{cont}, t_0)}\,.
\end{equation}
%%%%%%%%%%%%%%%%%%%%%%%%%%%%%%%%%%%%%%%%%%%%%%
In our case, it takes the form 
%%%%%%%%%%%%%%%%%%%%%%%%%%%%%%%%%%%%%%%%%%%%%%
%%%              Equation 6                %%%
%%%%%%%%%%%%%%%%%%%%%%%%%%%%%%%%%%%%%%%%%%%%%%
\begin{equation}
\Delta S(\lambda,t) = \frac{\Delta I_\mathrm{FOV}(\lambda,t)}{I^\mathrm{(N)}_\mathrm{FD}(\lambda_\mathrm{cont})}\,.
\end{equation}
%%%%%%%%%%%%%%%%%%%%%%%%%%%%%%%%%%%%%%%%%%%%%%
Integrating over this quantity over the interval $\Delta \lambda = \pm 0.5$\,\AA\ yields the differenced equivalent width $\Delta EW$ as 
%%%%%%%%%%%%%%%%%%%%%%%%%%%%%%%%%%%%%%%%%%%%%%
%%%              Equation 7                %%%
%%%%%%%%%%%%%%%%%%%%%%%%%%%%%%%%%%%%%%%%%%%%%%
\begin{equation}
\Delta EW = \int_{\lambda_0 - \Delta \lambda}^{\lambda_0 + \Delta \lambda} \Delta S(\lambda,t)\,\mathrm{ d}\lambda\,,
\label{eq:EW}
\end{equation}
%%%%%%%%%%%%%%%%%%%%%%%%%%%%%%%%%%%%%%%%%%%%%%
with $\lambda_0$ the wavelength of the spectra line in question. The $\Delta EW$ values are then scaled to facilitate comparison between lines with different widths and sensitivities to activity. The differenced equivalent width corresponds to the total change in the spectral line profile \citep{Otsu_2022}. For convenience, the relative values, denoted as $\delta EW$, have been adopted. They are normalized with respect to the reference value $|\Delta EW(t_0)|$ of particular line at the SST start time $t_0$ (Table~\ref{tab:flare_info}), such that $\delta EW(t) = \Delta EW(t) / |\Delta EW(t_0)|$. These normalized values are shown in Figs.~\ref{fig:X9.3}--\ref{fig:C1.2} alongside the ad hoc normalized soft \mbox{X-ray} flux measured by GOES in the $1\text{--}8$\,\AA\ passband. In some cases $\delta EW(t_0) = -1$ (see, e.g., Fig.~\ref{fig:X1.5}) causes the $\delta EW$ line to be separated from the other $\delta EW$ lines starting at 1. In these instances, as noted in the figure captions, an offset has been applied to the $\delta EW$ curve for improved clarity.

\subsection{Residual analysis via Voigt profile subtraction}

We analyze spectral residuals using a method similar to that of \citet{Ma_etal2024}, in which a Voigt profile is fitted to, and subtracted from, the contrast profiles (Eq.~\ref{eq:contrast_profile}). This approach has proven effective in isolating subtle features by removing the dominant central enhancement -- or, in some cases, central absorption -- visible in contrast profiles. The Voigt profile is defined as
%%%%%%%%%%%%%%%%%%%%%%%%%%%%%%%%%%%%%%%%%%%%%%
%%%              Equation 8                %%%
%%%%%%%%%%%%%%%%%%%%%%%%%%%%%%%%%%%%%%%%%%%%%%
\begin{equation}
V(\lambda,\sigma,\gamma )\equiv \int _{-\infty }^{\infty }G(\lambda',\sigma)L(\lambda-\lambda',\gamma )\,\mathrm{d}\lambda'\,,
\label{eq:Voigt}
\end{equation}
%%%%%%%%%%%%%%%%%%%%%%%%%%%%%%%%%%%%%%%%%%%%%%
where $G(\lambda,\sigma)$ and $L(\lambda,\gamma)$ denote the centered, area-normalized Gaussian and Lorentzian functions, respectively:
%%%%%%%%%%%%%%%%%%%%%%%%%%%%%%%%%%%%%%%%%%%%%%
%%%              Equation 9                %%%
%%%%%%%%%%%%%%%%%%%%%%%%%%%%%%%%%%%%%%%%%%%%%%
\begin{equation}
G(\lambda,\sigma) \equiv \frac{1}{\sqrt{2\pi}\,\sigma}\,\mathrm{e}^{-\frac{\lambda^2}{2\sigma^{2}}}\,,
\hspace{0.5cm}
L(x,\gamma) \equiv \frac {\gamma }{\pi (\gamma ^{2}+\lambda^{2})}\,.
\end{equation}
At each time frame $t$, the contrast profile $C(\lambda, t)$ is modeled by the function 
%%%%%%%%%%%%%%%%%%%%%%%%%%%%%%%%%%%%%%%%%%%%%%
%%%              Equation 10                %%%
%%%%%%%%%%%%%%%%%%%%%%%%%%%%%%%%%%%%%%%%%%%%%%
\begin{equation}
1 + \alpha(t) \, V(\lambda - \beta(t), \sigma(t), \gamma(t))\,, 
\label{eq:fit}
\end{equation}
%%%%%%%%%%%%%%%%%%%%%%%%%%%%%%%%%%%%%%%%%%%%%%
where $\alpha$, $\beta$, $\sigma$, and $\gamma$ are time-dependent fitting parameters. Subtracting the fit yields the residual spectrum:
%%%%%%%%%%%%%%%%%%%%%%%%%%%%%%%%%%%%%%%%%%%%%%
%%%              Equation 11               %%%
%%%%%%%%%%%%%%%%%%%%%%%%%%%%%%%%%%%%%%%%%%%%%%
\begin{equation}
R(\lambda, t) = C(\lambda, t) - 1 - \alpha(t)  V(\lambda - \beta(t), \sigma(t), \gamma(t))\,.
\label{eq:residuals}
\end{equation}
%%%%%%%%%%%%%%%%%%%%%%%%%%%%%%%%%%%%%%%%%%%%%%
Residuals $R(\lambda,t)$ are visualized using color maps, overlaid with contour lines of the fit, see Eq.~\ref{eq:fit}. Particular care was taken to ensure a robust fit to the core of each contrast profile. Initial guesses for the parameters $\alpha$, $\gamma$, and $\sigma$ were determined manually for the first frame of each event and subsequently propagated using a five-frame rolling average. Despite this, the fitting procedure was sensitive to initial conditions and did not always yield reliable results, for example, for the \Fe{i 6173}\ line in the Events~5 and 19  (Figs.~\ref{fig:X1.0} and \ref{fig:C1.2}, respectively). 

A small number of outlier frames occasionally disrupted the fitting sequence or introduced artificial features into the residuals. We therefore advise cautious interpretation of the residuals, particularly when they exhibit anomalous shapes. Nevertheless, the model generally captured the temporal evolution of the core of the contrast profile effectively.

%%%%%%%%%%%%%%%%
%%%  Fig 4   %%%
%%%%%%%%%%%%%%%%
%%%%%%%%%%%%%%%

\section{Results and discussion}
\label{sec:results}

This section presents a detailed analysis of \mbox{Sun-as-a-star} spectral signatures from 20 solar flares in 19 data sets, ranging in class from X9.3 to C1.2, observed by the SST between August 2011 and September 2024. We begin with an overview of unique spectral features imprinted in \mbox{Sun-as-a-star} spectra, which are specific to individual events, after which we discuss the individual events in Table~\ref{tab:flare_info}. Thereafter, we present a scale law between intensity and core intensity modulation. 

\subsection{Feature summary}

The contrast and residual profiles of the analyzed flares (Eqs.~\ref{eq:contrast_profile} and \ref{eq:residuals}) reveal global features shared by almost all events, along with a few less frequent yet distinct signatures that appear only in specific subsets of flares. From these profiles, we discern two sets of spectral features. Set~1 (Features~1--4) -- physical features that should be robustly detectable in disk-integrated observations with a high enough signal-to-noise-ratio. Set~2 (Features~5--6) -- spurious or misinterpreted features that imitate true detections but originate from the detection algorithm or from particular conditions of the flare. Particular attention is given to the detectability of CMEs in the stellar context.

\subsubsection{Physical features}

Our investigation finds the following features to originate from processes on the Sun, which can be interpreted physically. We distinguish them from other features.

\subsubsection*{Feature 1: Core intensity modulation}

Core intensity modulation is the most frequent feature of flare activity, detected across the full range of flare strengths. It manifests around the flare peak as the core intensity enhancement in all considered chromospheric lines except \HeD, where it appears as an absorption feature instead. The dark red contrast profiles in the top row of Figs.~\ref{fig:X9.3} and \ref{fig:X2.2} are a good example of the former, while the dark blue region at the center of the top right panel of Fig.~\ref{fig:M2.5}, exemplifies the latter. 
 
The specific flare behavior of the \HeD\ line is due to the photoionization-recombination mechanism populating the corresponding levels \citep{2019Libbrecht}. Consequently, the line initially appears in absorption before later transitioning into emission. The unambiguous variation in the intensity of the core is only absent in Event~8 (Fig~\ref{fig:M1.9}) most likely due to the small spatial overlap between the flare and the FOV (Table~\ref{tab:flare_info}, column 8). The corresponding residual profiles often but not always exhibit significant positive excess between wings of the contrast profile and the model function (e.g., Figs.~\ref{fig:X9.3} and \ref{fig:X2.2}), which suggests broadening of emission lines around the flare peak. This is characteristic for strong chromospheric lines.

\subsubsection*{Feature 2: Redshifted emission}

Several events display a transient, red‑shifted emission component that becomes most prominent at the flare peak and fades as the flare progresses. This signature is evident in Event~3 (Fig.~\ref{fig:X2.2}; \CaIR, \CaK), Event~5 (Fig.~\ref{fig:X1.0}; \CaIR), Event~6 (Fig.~\ref{fig:M3.0}; \CaIR), Event~7 (Fig.~\ref{fig:M2.5}; \Halpha, \Hbeta), and Event~17 (Fig.~\ref{fig:C2.0}; \CaIR, \CaK). We attribute this temporary red shift to down‑flowing, cool, condensed plasma that drains along flare loops into the flare ribbons, which are largely unobscured by overlying filaments (see the corresponding animations$^{\ref{foot:git}}$). This phenomenon is commonly referred to as coronal rain  \citep[see e.g., ][]{2024PietrowRibbon}. All of these events lie at heliocentric directional cosines of $\mu\approx 0.93-0.64$ (Table~\ref{tab:flare_info}), a range that maximizes the line‑of‑sight projection of the flow velocity and thereby enhances the red‑shifted components. This interpretation is in line with the findings of \citet{Yu2025}. A redshifted emission -- referred to as red asymmetry -- is evident in the \Halpha\ profile associated with a superflare on YZ Canis Minoris, reported by \citet[][panel~(d) in Fig.~3]{Namizaki2023}.

 \subsubsection*{Feature 3: Imprints of filaments and flare loops}

A comparatively rare yet distinctive signature is observed in Events~8, 11, and 18 (Figs.~\ref{fig:M1.9}, \ref{fig:C8.4}, and~\ref{fig:C1.5}). It appears as a blue absorption feature in the contrast profiles and residuals and can be present in the blue and sometimes the red wings of these profiles. Inspection of the associated \Halpha\ animations$^{\ref{foot:git}}$ indicates that this behavior arises when dynamic structures  (flare loops in Event~8 and rising filaments in Events~11 and~18) partially obscure the underlying flare ribbons. Remarkably, these spectral signatures resemble those associated with an eruptive filament from a superflare on a solar-type star and a C5.1-class solar flare and filament eruption \citep[][see panels c and d in Fig.~1, and panel b in Fig.~2, respectively]{Namekata2021}.

\subsubsection*{Feature 4: Temporal offset between $\delta EW$ and GOES peaks}

Feature~1 is accompanied by an enhancement in the normalized differential equivalent width ($\delta EW$) of the contrast profiles, as illustrated in the upper right panels of Figs.~\ref{fig:X9.3}--\ref{fig:C1.2}. In Events~7, 10, 15, 17, and 18 (Figs.~\ref{fig:M2.5}, \ref{fig:M1.1}, \ref{fig:C1.9_and_C2.4}, \ref{fig:C2.0}, and \ref{fig:C1.5}, respectively), the $\delta EW$ enhancement appears to precede the peak in the GOES data. In contrast, for Events~1, 3, and 6 (Figs.~\ref{fig:X9.3}, \ref{fig:X2.2}, and \ref{fig:M3.0}, respectively), the peak times of $\delta EW$ and the GOES data appear to coincide. Only in Event~9 (Fig.~\ref{fig:M1.8}) does the $\delta EW$ peak clearly lag behind the GOES peak. A definitive assessment of the temporal relationship between the $\delta EW$ and GOES peak times will be presented in a separate dedicated investigation. For the remaining events, incomplete temporal coverage precludes a reliable evaluation of this relationship.

\subsubsection{Spurious features}

We attribute these features primarily to instrumental and seeing-induced effects, which should be carefully considered in future studies involving a limited FOV. We highlight these features as they can be misinterpreted as physical features.

\subsubsection*{Feature 5: False quasi-periodic oscillations}

A closer inspection of the residual profiles for Events~1--7 and 17--19 (Figs.~\ref{fig:X9.3} -- \ref{fig:M2.5} and \ref{fig:C2.0} -- \ref{fig:C1.2}, respectively) reveals quasi-periodic patterns in the contrast profiles and residuals (appearing as alternating blue and red) that are apparent in some or all of the spectral lines. An investigation into their origin shows that they are either caused by seeing-induced, quasi-periodic smearing of bright flare features (e.g., Event~1), or by minor tracking issues that result in parts of the flare intermittently drifting in and out of the field of view during the time series (e.g., Event~2). These effects reflect limitations inherent to observations with a restricted field of view and are not expected to occur in stellar observations, where spatial resolution is not a factor.

\subsubsection*{Feature 6: False profile modulation and abrupt jumps}

\begin{figure*}
\centering
\includegraphics[width=0.75\textwidth]{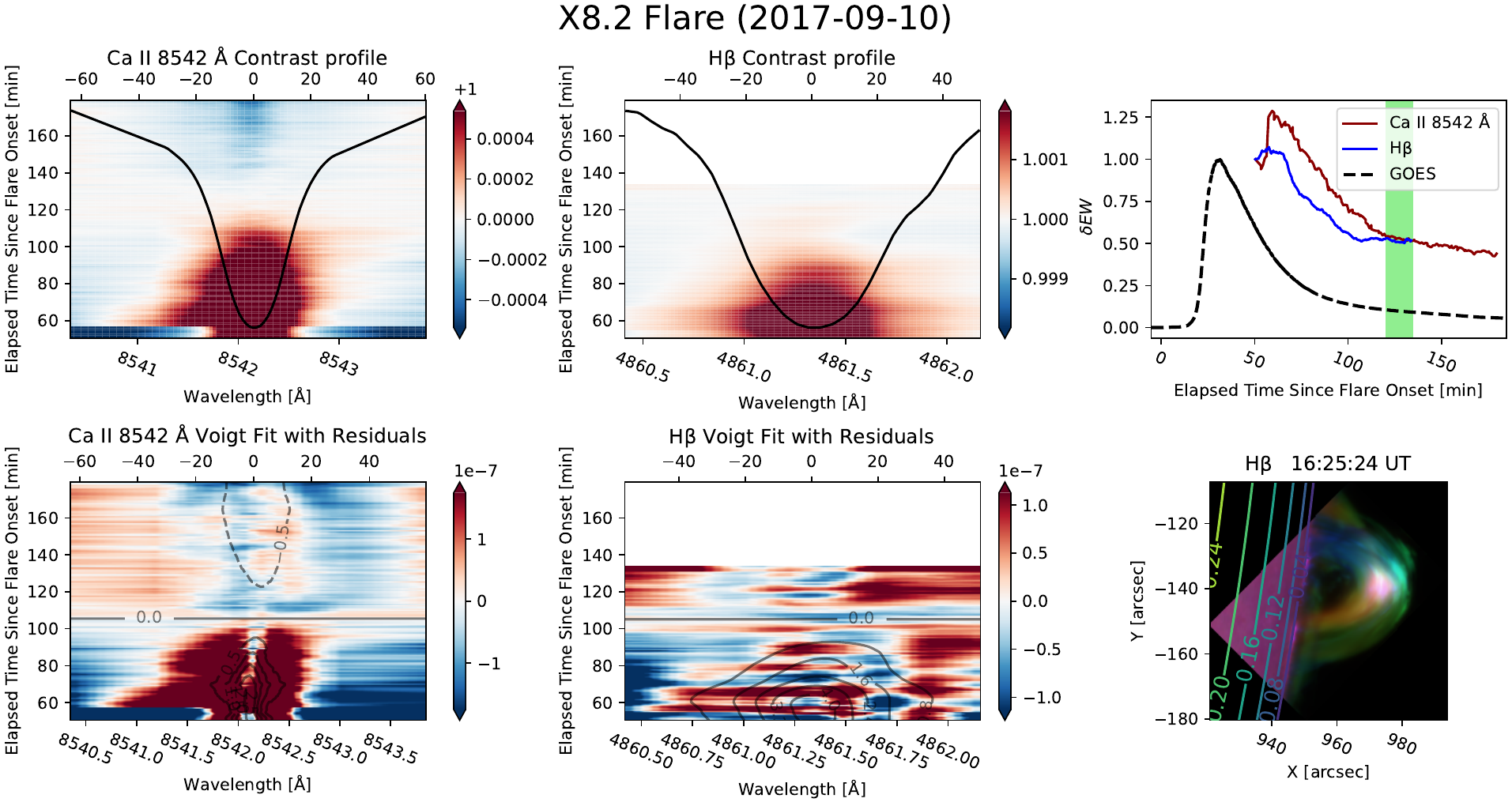}
\caption{Event~2. Same as Fig.~\ref{fig:X9.3} but for \CaIR\ and \Hbeta\ disregarding \Fe{i 6302} and \CaHK. In the top left panel, the colorbar values of $C(\lambda, t)$ are offset by +1, as indicated at the top of the colorbar. This offset is also applied in some of the subsequent figures.}
\label{fig:X8.2}
\end{figure*}

\begin{figure*}
\centering
\includegraphics[width=\textwidth]{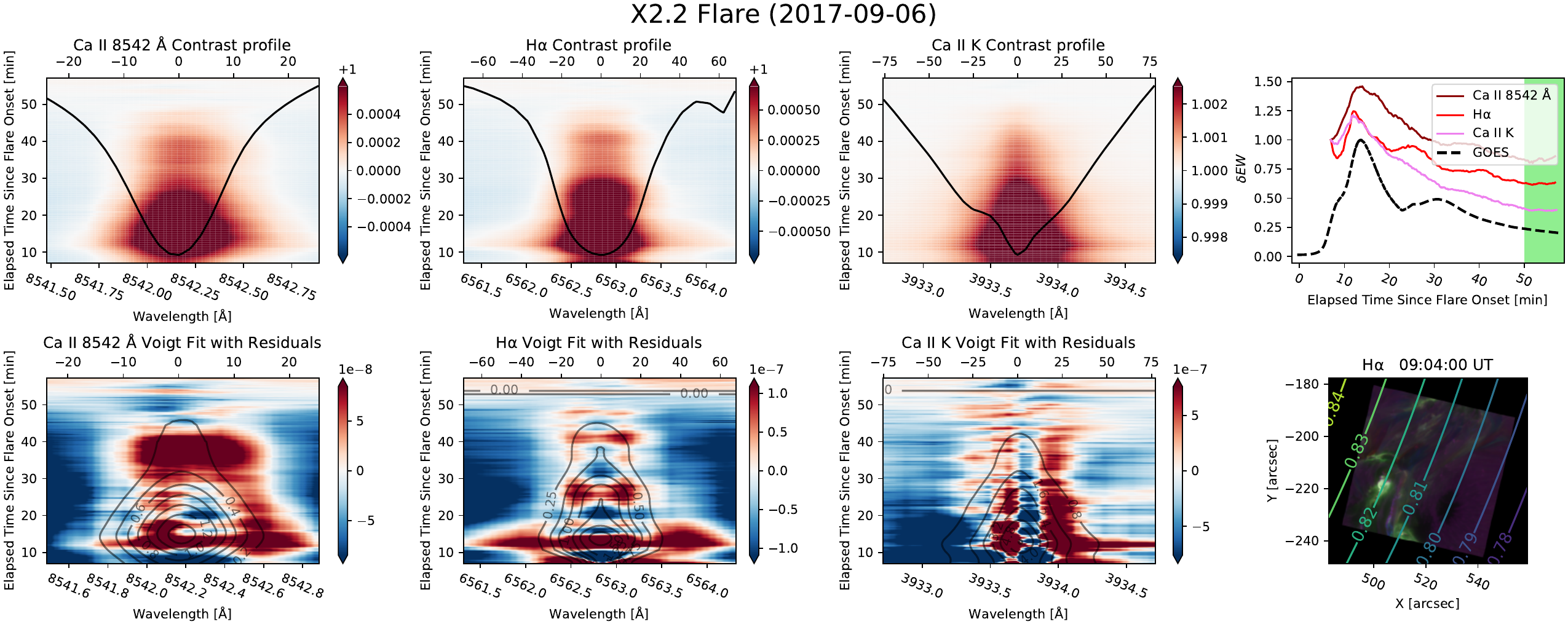}
\caption{Event~3. Same as Fig.~\ref{fig:X9.3}. }
\label{fig:X2.2}
\end{figure*}

Events~2, 7, 8, 9, and~15 (Figs.~\ref{fig:X8.2}, \ref{fig:M2.5},
\ref{fig:M1.9}, \ref{fig:M1.8}, and~\ref{fig:C1.9_and_C2.4},
respectively) exhibit brief anomalies in their contrast profiles at the very start of their
time series'. This anomaly is a processing artifact -- an unrealistically weakened line
core, thus rendered in dark blue. It arises because the reference interval
used for normalization -- indicated by the green bars in the upper-right
panels of respective figures -- was chosen after the flare peak, combined
with the particular shape of the reference line profiles. In addition to this initial spurious modulation, the residuals of
Events~2 and~9 (Figs.~\ref{fig:X8.2} and~\ref{fig:M1.8}) show abrupt
jumps caused by small telescope repointing undertaken to
keep the flare optimally centered within the field of view.

\subsubsection{Detectability of CMEs}

% prologue, context
In the context of the rapidly expanding literature on stellar CMEs \citep[e.g.,][]{Veronig_etal2021,Namekata2021,Leitzinger_etal2022,Leitzinger_etal2024,LeitzingerOdert2022}, this subsection focuses on identifying physical CME signatures in Sun-as-a-star observations and spurious, CME-like features. Distinguishing between the two is crucial for robust stellar-CME detection.

% CME-positive flare statistics
An inspection of the \href{https://cdaw.gsfc.nasa.gov/CME_list/}{SoHO/LASCO CME Catalog} \citep{Gopalswamy_etal2009} revealed ten events (1--5, 7, 9--11, and 17 in Table~\ref{tab:flare_info}). Thus, 53\% of the flares in our sample produced a cataloged CME. Among these events associated with CMEs, only Event~11 (Fig.~\ref{fig:C8.4}) exhibits the typically associated contrast and residual profiles (a blue-wing absorption feature, occasionally mirrored in the red wing) despite its associated CME being the second weakest of the set. Within the SST spectral window, these signatures closely resemble those of CME events~(5)--(7) in \citet[][Table~1; panels~(a) of Figs.~10, 13, 15]{Otsu_2022}, all of which followed filament eruptions. While CMEs can occur without an accompanying flare it is believed that strong solar flares typically have associated CMEs \citep[e.g.,][]{Yashiro2009}. However, our data reveals no common spectral fingerprint shared by all CME-associated events. On top of that, all but one of these events showed no sign of the classical CME-Doppler signature in the chromospheric spectral lines. However, we caution that this could be a result of the limited FOV of the SST, which simply does not capture the full region.

% CME-negative flare statistics
Conversely, nine flares (6, 8, 12--16, 18, 19) show no cataloged CME (Table~\ref{tab:flare_info}), underscoring that flare occurrence alone is an unreliable CME predictor. Moreover, Doppler signatures very similar to those often associated with CMEs have been observed in Events~8 and 18 (Figs.~\ref{fig:M1.9} and \ref{fig:C1.5}). In both cases, gravitationally bound up-flows cause the Doppler signatures. These false positives, along with the false negatives from flares with associated CMEs, warrant further research into the accuracy of the association between CMEs and the corresponding Doppler features.

%%%%%%%%%%%%%%%%
%%%  Fig 5   %%%
%%%%%%%%%%%%%%%%
%%%%%%%%%%%%%%%%%

%%%%%%%%%%%%%%%
%%%  Fig 6  %%%
%%%%%%%%%%%%%%%
%%%%%%%%%%%%%%%%%

\subsection{Individual event summaries}

In this section, we discuss each event in detail and list the causes for variations in the \mbox{disk-integrated} contrast profiles, residual profiles, and $\delta EW$s derived from NESSI. These are shown together with a representative image of the flare and lines showing its location on the disk. Each subheading is listed between Event~1 to 19. The overall findings of this section were listed in the Feature summary above and are recapped in the Conclusions.

Observations made in the \HeD\ and iron lines are shown in a different colorbar than the rest to differentiate them from the optically thick chromospheric lines. In the case of \HeD\ there is also no limb darkening data available to process the lines with NESSI. As a result, we present a simple \mbox{FOV-integrated} profile. 

Each flare is accompanied by its GOES and $\delta EW$s curves, to better illustrate potential delays between the different lines. 

Each FOV is displayed as a COlor-COlapsed PLOTs \citep[COCOPLOTS;][]{druett_cocoplot_2022}, which encode spectral information into color by multiplying the spectra with three offset Gaussians which are then loaded into the R, G, and B channels of the image. We refer the reader to the paper for more examples of this method. 
Videos of the events are provided in the supplementary materials, showing a time  series of the flare in each spectral line, along with the \mbox{disk-integrated} profile.

\subsubsection*{Event 1: An X9.3 flare on 2017 September 6}

An X9.3 flare occurred in the active region NOAA 12673 on 2017-09-06 at $\sim$11:53~UT. Approximately 40\% of the flare ribbon area was captured in the FOV of the SST, and therefore, we scaled the recorded flare emission by a factor of 2.5 before passing it onto NESSI as described in Section~\ref{sec:contrast}. The average $\mu$ value for the flare is 0.82. The flare was associated with a Halo CME with a recorded velocity of 1571~\kms\ and \mbox{CME-driven} shock that intercepted the Earth causing a significant space weather event \citep[e.g.,][]{2024Yu_cme_x93_flare}. This flare was first described in \citep{2019Quinn_Sunquake}.

Fig.~\ref{fig:X9.3} shows contrast profiles for the spectral lines in the X9.3 flare, with contrast taken relative to the time frames shown with a green bar in the $\delta EW$ panel. Note that a zero residual does not indicate a flat-line profile, rather no difference to the reference profile used. The \CaK\ line appears to have a significantly different contrast level with a much higher sensitivity to small-scale effects.

The shape of all three contrast profiles are similar with \CaIR\ showing the strongest red asymmetry, showing the increased sensitivity to Doppler shift at high wavelengths, given $\Delta\lambda = v_0\lambda_0/c$. A clear temporal offset between the $\delta EW$ peaks can be seen. 

In the residual profiles, more structure can be seen. In \Halpha\ we see an emissive (red) blueshifted emission lasting roughly 15 minutes, as well as a longer-lasting redshifted emission for about 30 minutes, which weakens considerably after 20. The former coincides with the formation of the flare arcade which has material in emission in these lines moving up and down its loops, while the latter also coincides with the formation of the flare ribbon which compresses the atmosphere causing a redshift until it expands out of the FOV around 30 minutes. In \CaK\ we see a similar emissive blue, near-wing enhancement that lasts for 30 minutes in a reasonably stable state. The emission from redshifted material has high velocities early in the flare, and slowly decreases with a similar timescale to the slowing down of the ribbon expansion. After the 30 minute mark, a faint signal showing increasing emissive blue and red shifts can be seen which seems to be connected to flows connected to areas outside of the FOV. We believe that the pulsations are not associated with true quasi-periodic pulsations, but are due to variations in seeing conditions which average out small areas with high Doppler shifts, and move bright regions near the edge of the image in and out of the FOV.

\subsubsection*{Event 2: An X8.2 flare on 2017 September 10}

An X8.2 flare occurred in the active region NOAA 12673 on 2017-09-10 at $\sim$16:30~UT. Approximately 50\% of the flare ribbon area was captured in the FOV of the SST. The average $\mu$ value for the flare is 0 as the flare occurred very close to the limb and the associated emission comes primarily from the arcade that extends beyond it (see Fig.~\ref{fig:X8.2}. lower right panel). A very strong halo-CME of 3163~\kms\ is associated with this event. This flare was first described by \citet{kuridze2020spectral}.

The contrast profiles and $\delta EW$s show a very stable increase in emission, which is visible for longer in \CaIR\ than \Hbeta\ likely due to the higher contrast in emissivity caused by the flare in these lines \citep[e.g.,][]{Capparelli2017}. Additionally, a sharp jump can be seen in the \Hbeta\ profiles early in the time series, this is due to a change made in the observing program where the \CaHK\ lines were removed from the observational cycle to increase the cadence of \Hbeta. During this change, the FOV jumped and excluded part of the flare arcade. 

The \Hbeta\ residuals show a decreasing emissive blueshift and enhancement, and an increase in red-wing emission. This is an observational effect, as the arcade is slowly moving out of the FOV due to improper tracking, with the drift over time gradually cutting off the part of the arcade flows with blueshifts, while not affecting the red part. A similar trend can be seen in the contrast profiles of \CaIR. If not for these effects, we would expect a very stable increase over the flare's lifetime. 

\subsubsection*{Event 3: An X2.2 flare on 2017 September 6}

An X2.2 flare occurred in the active region NOAA 12673 on 2017-09-06 at $\sim$09:00~UT, three hours before the X9.3 flare discussed in a previous section \citep[for discussion of the relationship between these flares see][]{Liu2018_X2.2_X9.3, 2024pietrowHarps}. Approximately 50\% of the flare ribbon area was captured in the FOV of the SST. The average $\mu$ value of the flare is 0.8. The SST data associated with this flare was first published by \citet{Vissers2021}. A \mbox{slow-moving} CME of 391~\kms\ is associated with this event.

The GOES curve shows that this is a complicated event with the impulsive increase slowing significantly at around 10 minutes before increasing strongly again and reaching the peak at 15 minutes. A subsequent peak can also be seen just beyond 30 minutes, suggesting that several bursts of reconnection may have occurred. The behavior in the $\delta EW$ of the \Halpha\ line seems to track the GOES evolution, with the second peak around 25 minutes, and the third around 40 minutes. However, when comparing to SDO, the secondary GOES peak appears to happen outside of the FOV. The peak at around 25 minutes in \Halpha\ is due to the disappearance of a clump of evaporating material that hung partially over the flare ribbons, absorbing some of this emission, similar to Event~1 in \citet{Otsu_2022}. When it disappeared, more of the light passed through. Such absorption features are visible in the wings of the residual profiles, seen as dark blue patches in the far wings from 20 to 30 minutes. The peak at 40 minutes seems to be due to poor seeing. These examples highlight that overlying cool material overlapping the flare ribbons in the line of sight can cause significant apparent Doppler shifts in the net flare emission, without any strong signal present from within the ribbons themselves. Such emission should therefore be treated with caution and not necessarily taken to imply the presence of a CME.

\subsubsection*{Event 4: An X1.5 flare on 2014 June 10}

An X1.5 flare occurred in the flare-prolific active region NOAA~12087 on 2014-06-10 at $\sim$12:50~UT. The flare was associated with a large 925~\kms\ CME. Approximately $75\%$ of the flare ribbon area was captured in the FOV of the SST. The flare is located very close to the solar limb, and the average $\mu$ value of the flare is $0.14$. 

\begin{figure}
\centering
\includegraphics[width=\linewidth]{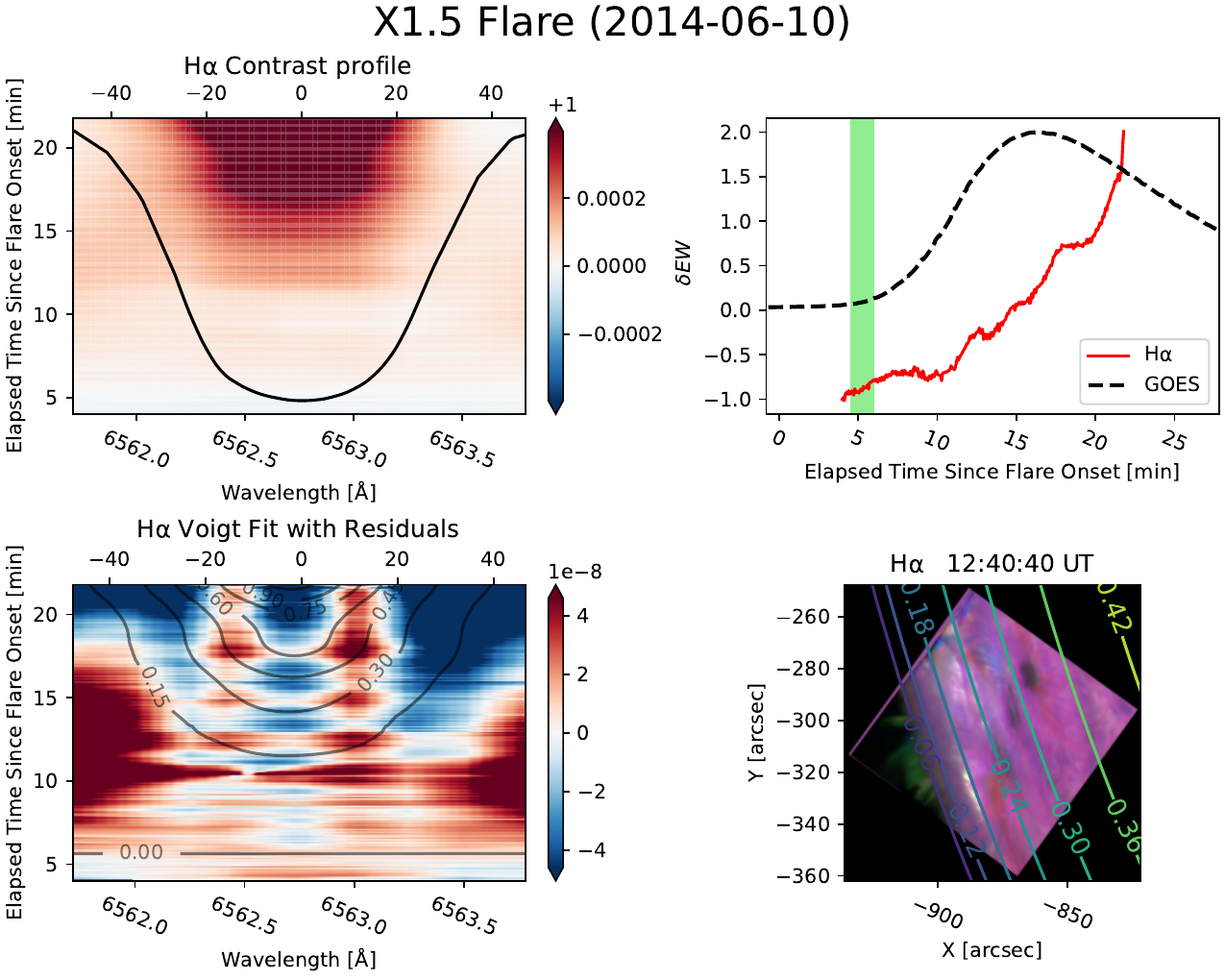}
\caption{Event~4. Same as Fig.~\ref{fig:X9.3} but for \Halpha. }
\label{fig:X1.5}
\end{figure}

The flare ribbons themselves produce strong broad enhancements in the net \Halpha\ emission, seen as positive (red) contrast across the wavelength range in the Voigt residuals (see Fig.~\ref{fig:X1.5}). In particular, this enhancement is noticeable in the wings of the profiles, seen as enhancements at times of 8 to 16 minutes. Coronal rain showers manifest in the \Halpha\ line spectra later in the observational sequence, which appear to manifest in the negative (blue) Voigt residuals at times from 18 minutes onward. Again, the pulsations in intensity appear to relate to changes in seeing conditions rather than any \mbox{quasi-periodic} flare processes.

%%%%%%%%%%%%%%%%
%%%  Fig 7   %%%
%%%%%%%%%%%%%%%%
%%%%%%%%%%%%%%%%

%%%%%%%%%%%%%%%%
%%%  Fig 8   %%%
%%%%%%%%%%%%%%%%
\begin{figure*}
\centering
\includegraphics[width=0.75\textwidth]{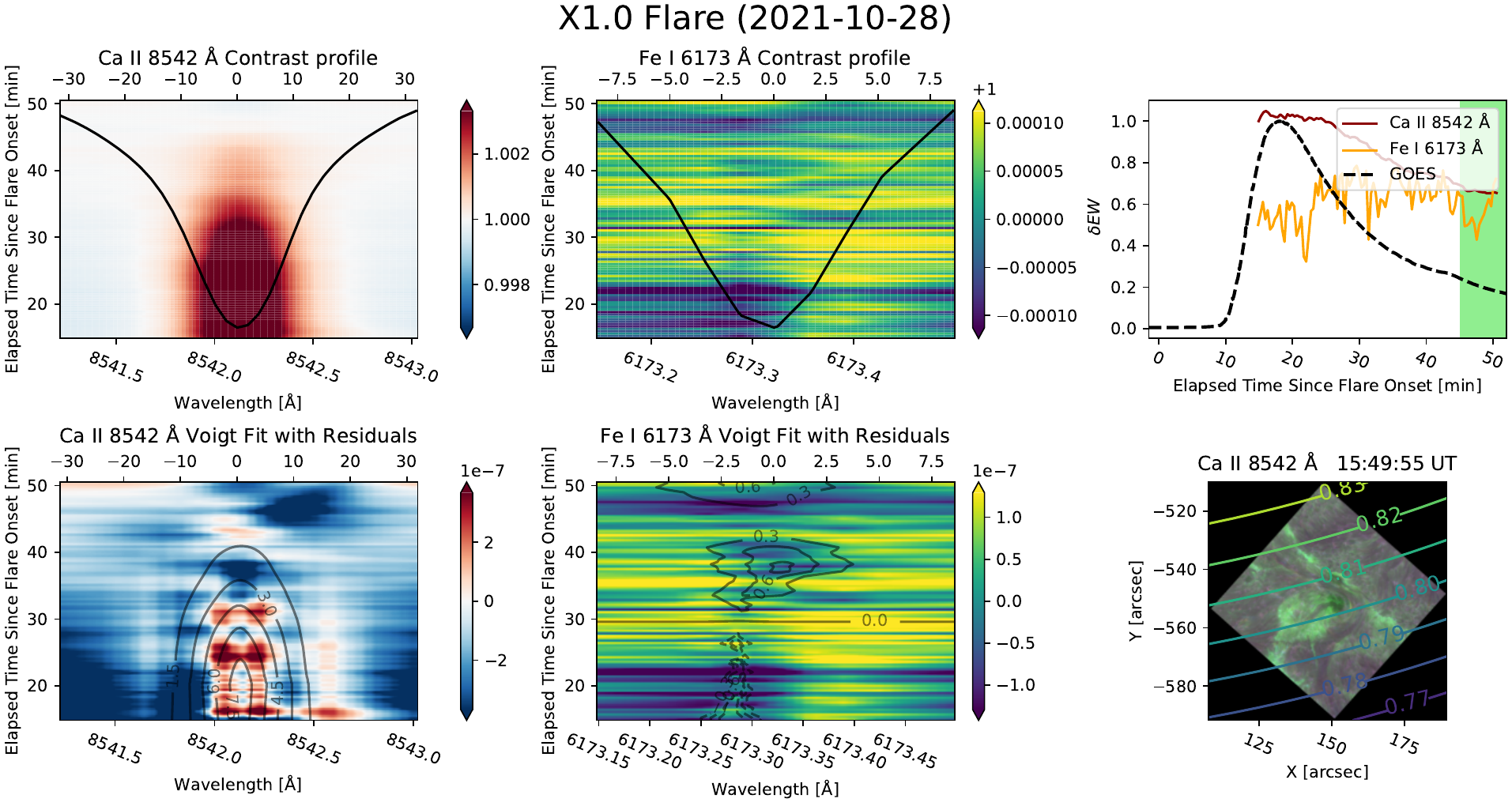}
\caption{Event~5. Same as Fig.~\ref{fig:X9.3} but for \CaIR\ and \Fe{i 6173}. The $\delta EW$ of \Fe{i 6173} has been offset by 1.5 for better comparison.}
\label{fig:X1.0}
\end{figure*}
%%%%%%%%%%%%%%%%

\subsubsection*{Event 5: An X1.0 flare on 2021 October 28}

An X1.0 class flaring event occurred on 2021 October 28 around 15:17~UT. The flare happened in the active region NOAA 12887 with a $\mu$-value 0.804 and was observed in the \CaIR\ and \Fe{i 6173} spectral lines with a 21.6-s cadence. Approximately $10\%$  of the flare ribbon area was captured in the FOV of the SST. The flare was associated with a large 1519~\kms\ CME, although this occurred far from the FOV pointing.

This flare occurred after a filament eruption and consisted of two main parallel flare ribbons, with some smaller flaring ribbons around them. These later merged into a composite ribbon \citep{Yamasaki2022,Guo23}. The SST FOV centered on a rather dynamic part of the flare (Fig.~\ref{fig:X1.0}), as the southern ribbon sweeps across this area at minute 11 (before the start of the SST observation) and then disconnects from the main ribbon at minute 16 (around the start of the SST observation) and starts dying down. The flare ribbon has two further small expansions around minutes 25 and 30, after which the emission (Fig.~\ref{fig:X1.0}, top right panel) rapidly decays and small upflows and coronal rain become visible. The enhancements to emission on the red and blue sides of the line core are due to line broadening caused by the flare heating the lower atmosphere, and the expanding front of the flare ribbon causes red-shifted emission enhancements seen in the Voigt residuals at wavelengths around 8542.6\,\AA\ (Fig.~\ref{fig:X1.0}, bottom left panel). 

In the iron line, some red-shifted emission can be seen starting at 25 minutes which can be attributed to strong flows in the rapidly evolving light bridge crossing the spot. 

\subsubsection*{Event 6: An M3.0 flare on 2022 May 20}

An M3.0 flare occurred in the active region NOAA 13014 on 2022-05-20 at $\sim$07:35~UT. At the first frame of observation, almost full capture of the ribbon area at that time is achieved. However, in the evolution of the flare the ribbons expands outwards partially leaving the FOV. We have approximated this by setting overall 90\% ribbon area capture. The flare is near the center of the solar disk, and the average $\mu$ value of the flare is 0.927. No CME is associated with this event. 

\begin{figure}
\centering
\includegraphics[width=\linewidth]{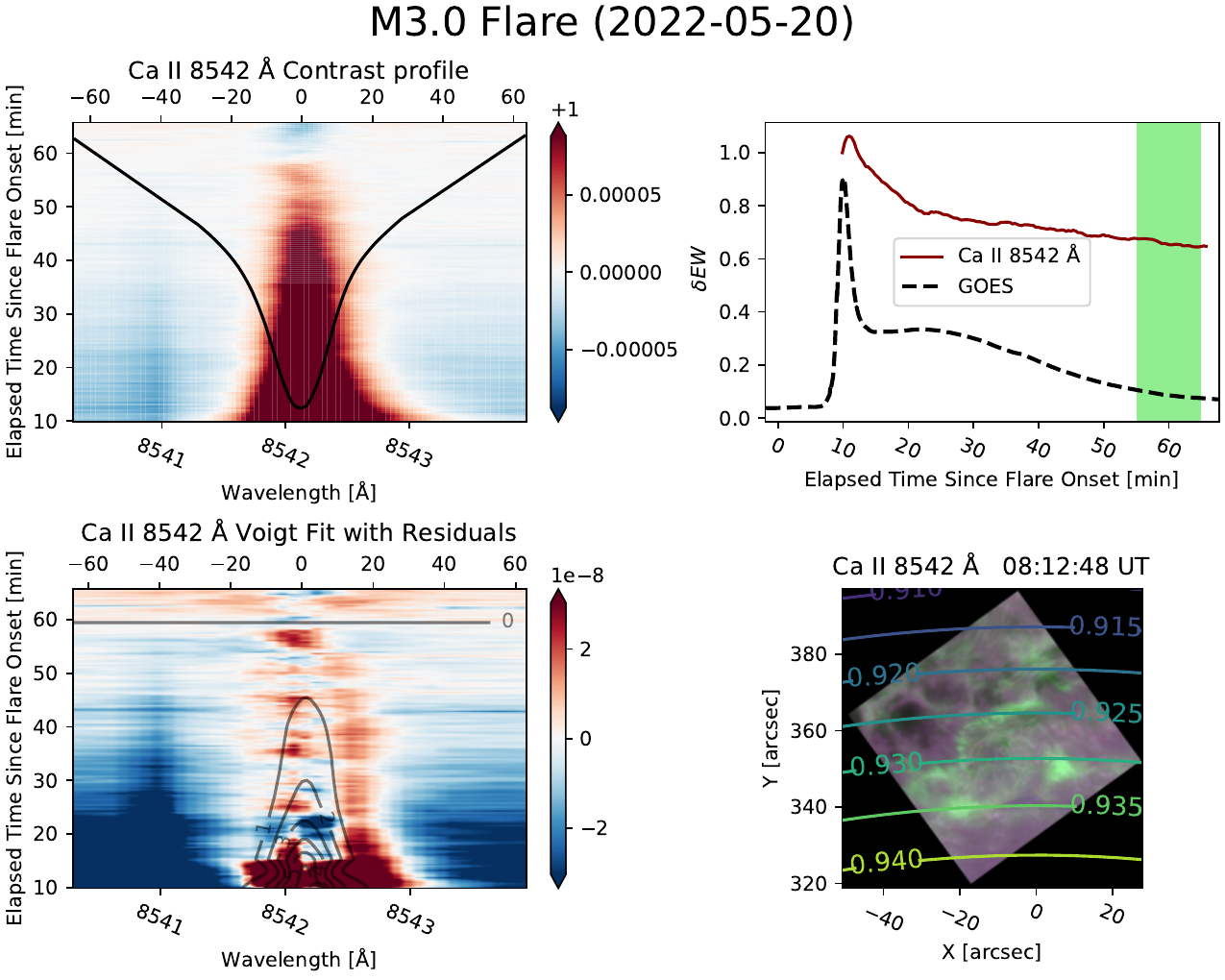}
\caption{Event~6. Same as Fig.~\ref{fig:X9.3} but for \CaIR.}
    \label{fig:M3.0}
    \end{figure}

The observation starts at the peak of the flare in the GOES signal and a regular decay in the line enhancement is seen from then onward, both in the contrast profile, Voigt contour lines, and in the normalized differenced equivalent width (See Fig.~\ref{fig:M3.0}). The GOES curve has a secondary plateau which is only visible in the residuals of the Voigt fit. At the peak of the flare, the ribbon is concentrated and shows strong line center enhancements on the flare ribbon, this produces a central reversal in the line core of the net (NESSI) emission profile. The amount of central reversal decreases between 10 to 20 minutes, causing strong changes to the fitting parameters for the Voigt residuals. Thereafter, the profile contains a double-peaked reversal similar to that of the \CaHK\ lines, and this is consistently redshifted throughout the flare's lifetime.
There is also some blueshifted material evident in the video COCOPLOTS during the time span 10--20 minutes and these appear to be associated with the contrast profile and Voigt residuals showing a strong negative value at around 8541\,\AA. We note that this flare has no associated CME. This flare has a very similar contrast profile to the 1D simulations for flares close to the disk center shown in \citet{Yu2025}.

%%%%%%%%%%%%%%%%
%%%  Fig 9   %%%
%%%%%%%%%%%%%%%%
%%%%%%%%%%%%%%%%

%%%%%%%%%%%%%%%%
%%%  Fig 10   %%%
%%%%%%%%%%%%%%%%
\begin{figure*}
\centering
\includegraphics[width=\textwidth]{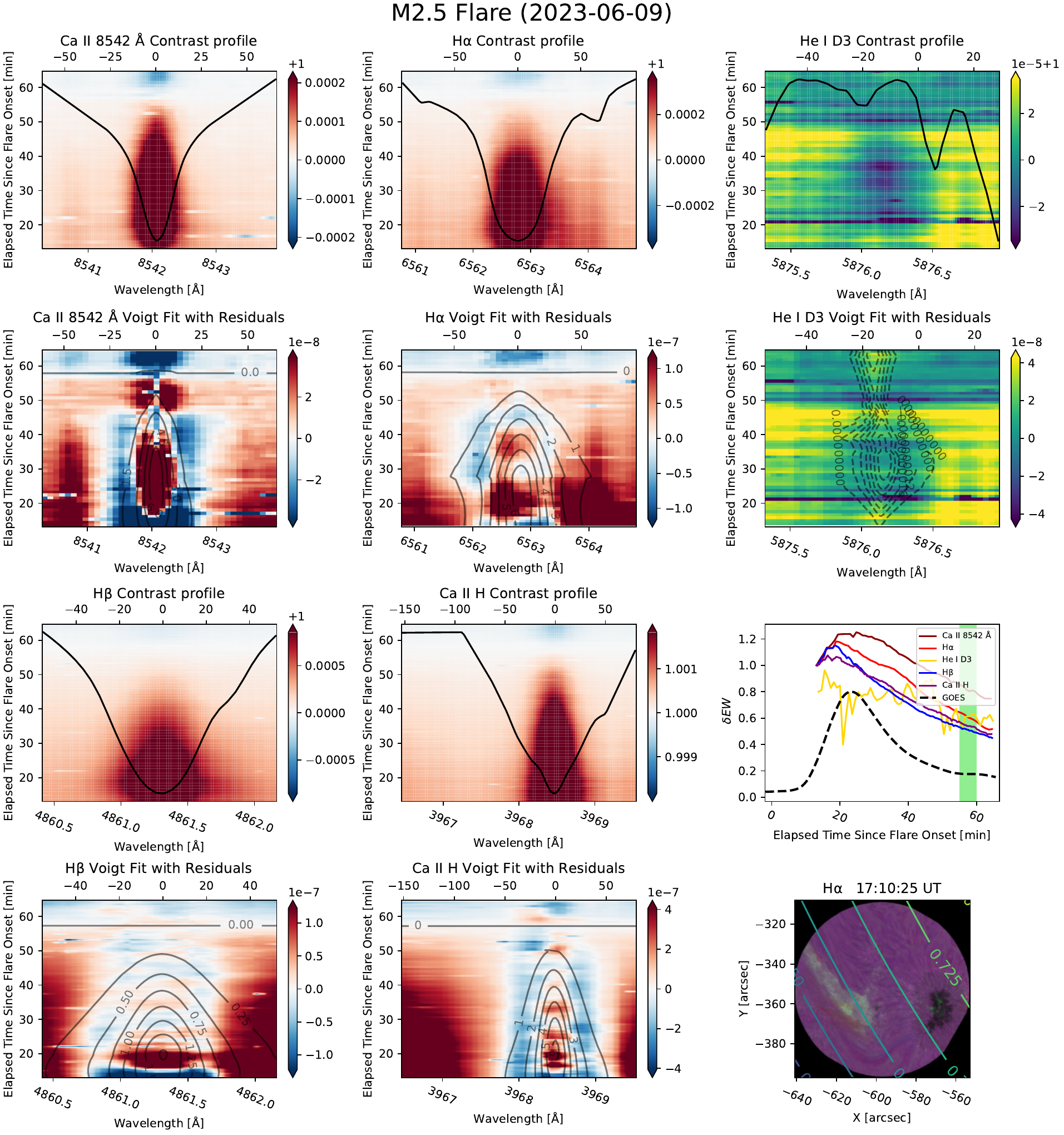}
\caption{Event~7. Same as Fig.~\ref{fig:X9.3} but for \CaIR, \Halpha, \HeD, \Hbeta, and \CaH. In the top right panel,the absorption features in the \mbox{quiet-Sun} profile (black curve) are blends of multiple telluric H$_2$O and solar metallic lines (see Fig.~2 in \citet{Libbrechtetal2017} and the top panels of Fig.~4.4 in \citet{Libbrecht2019}). The \HeD\ contrast profile corresponds to the dark region approximately within the ranges $5876.00-5876.50$\,\AA\ and $25-45$\,min. The colorbar values of $C(\lambda, t)$ are on the order of $10^{-5}$ and offset by +1, as indicated at the top of the colorbar as $1\mathrm{e}-5 + 1$. This formatting is also used in some of the subsequent figures. In the right column, third panel from the top, the $\delta EW$ curve of \HeD\ has been offset by 1.8 for better comparison.}
\label{fig:M2.5}
\end{figure*}
%%%%%%%%%%%%%%%%%

%%%%%%%%%%%%%%%%
%%%  Fig 11  %%%
%%%%%%%%%%%%%%%%
\begin{figure*}
\centering
\includegraphics[width=0.75\textwidth]{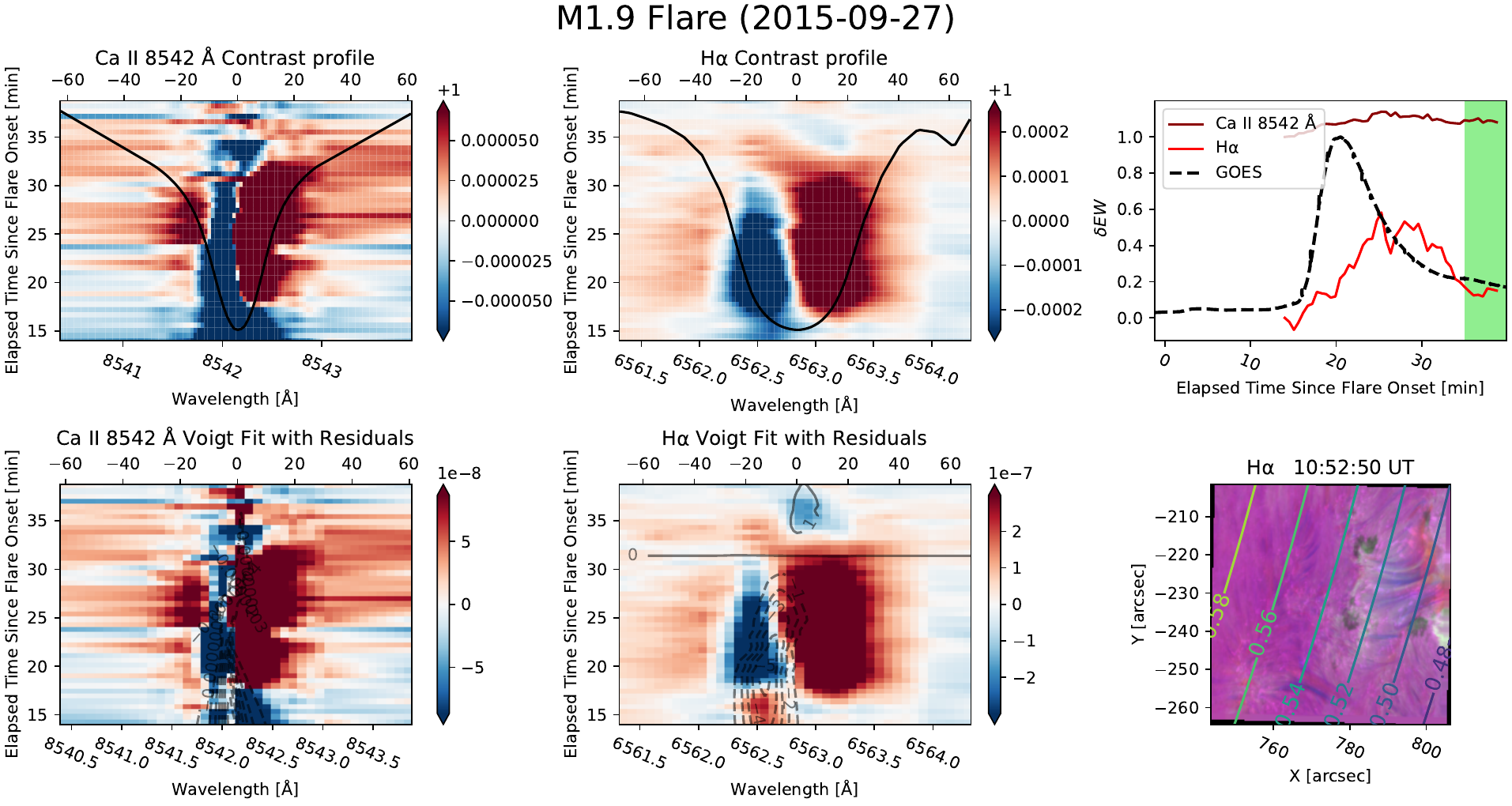}
\caption{Event~8. Same as Fig.~\ref{fig:X9.3} but for \CaIR\ and \Halpha. The $\delta EW$ of \Halpha\ has been offset by 1 for better comparison.}
\label{fig:M1.9}
\end{figure*}
%%%%%%%%%%%%%%%%%

\subsubsection*{Event 7: An M2.5 flare on 2023 June 9}

An eruptive M2.5 flare occurred in the active region NOAA 13331 on 2023-06-09 at $\sim$16:48~UT. Approximately 40\% of the flare ribbon area was captured in the FOV of the SST/CRISP (\CaIR, \Halpha, and \HeD), with slightly less over the SST/CHROMIS (\CaH\ and \Hbeta) FOV. The average $\mu$ value of the flare is 0.688. The event has an associated CME with 651~\kms. The analysis is displayed in Fig.~\ref{fig:M2.5}. 

The times of maximum contrast profile enhancements in different optically thick chromospheric lines are again offset. The peak occurs earliest in \CaK\ (typically forms the highest in the atmosphere) and then \Halpha\ and \Hbeta, and the latest peak is seen in \CaIR\ (typically the lowest forming of these lines). The contrast in the Helium line is extremely low, the low level of line core absorption that is present in the contrast profiles peaks tens of minutes after all other lines. In the resolved images as the flare ribbon expands we see dimming, follow by some brief brightening, and a prolonged period of darkened absorption in the line center. This is in general agreement with the theoretical description provided by \citet{Kerr_2021_He_dimming}. 

Viewing the spatially resolved spectral data, we note that in the \Halpha, \Hbeta, and \CaH\ images and profiles there is strong emission at the start of the flare in the red wing as the flare ribbon forms and compresses the chromosphere, as in \citet{2017Druett} and \citet{2024PietrowRibbon}. While for \Halpha\ this emission covers wavelengths where there is a telluric line (6564 \AA), the resolved signal is clearly dominated by the flare ribbon.

A fast-moving upflowing jet is visible from near the sunspot ($-570$\arcsec, $-370$\arcsec) from 30 minutes up to about 50 minutes, which absorbs the underlying emission. This jet is also visible in the resolved images of the other lines, but not clear in the contrast profiles or Voigt fits. It is possibly merged with the Voigt fit residuals found around the core emission. In helium the flare ribbon is visible as absorption, and most clear around 20 minutes when the ribbon forms, and at roughly 28 minutes when a secondary ribbon appears. 

The streak of positive values in the residual fits for the \CaIR\ line at 8540.8~\AA\ is due to a combination of unequal wavelength sampling and the drift of the FOV of the instrument over time, bringing more of the penumbra into view with time. This artifact would not be present in true \mbox{Sun-as-a-star} observations.

There is a blue-shifted absorption feature in the \CaIR\ and \Halpha\ line cores at times after $t\sim60$ minutes. This is a result of taking the reference frames for the contrast profile around 55-60 minutes during the gradual phase of the flare. As the ribbon activity decreases slowly, the "contrast" therefore appears negative in the core at later times.

\subsubsection*{Event 8: An M1.9 flare on 2015 September 27}

An M1.9 flare occurred in the active region NOAA 12423 on 2015-09-27 at $\sim$10:40~UT. Approximately 10\% of the flare ribbon area was captured in the FOV of the SST. The average $\mu$ value of the flare is 0.5. No CME is associated with the event. This flare was first described in \citep{Kuridze2018} and was observed during the IRIS and SST coordinated campaigns described in \citet{Luc2020}.

In this instance throughout the observations, there is an arcade of loops that overlaps the flare ribbon emission in the line of sight (Fig.~\ref{fig:M1.9}, bottom right panel). From the line profiles and the cocoplot (particularly in \CaIR), one can infer there is a transfer of relatively cool material along these arched fibrils occurring predominantly from one ribbon at ($X, Y$) = (785\arcsec, $-200$\arcsec) toward the other ribbon at ($X, Y$) = (780\arcsec, $-225$\arcsec) producing a net emission showing a blue Doppler shift in the line core. 

As a result, the contrast profiles are strongly impacted throughout the duration of this action. The line core minimum position shifts left, to the shorter or bluer wavelengths.
Therefore, relative to the reference spectrum, the emission in the wing wavelength left of the center appears lower, and that to the right appears greater. Additionally, the peak in the line $\delta EW$s is much later than the GOES peak in both cases. We believe that this is due to the cool material covering the flare ribbons. 

When looking at a contrast profile or residual, a persistent net Doppler shift to one side generates  a pair of enhanced and decreased emission on opposite side of the line core. Which side is enhanced and which decreased depends on whether the relevant material is behaving as an emission line (enhancement matches Doppler shift) or an absorption line (enhancement in opposite direction to Doppler shift). However, this signal is not associated with a related CME, despite exhibiting the CME spectral signature that stellar observers often look for \citep[e.g.,][]{Namekata2021}.

%%%%%%%%%%%%%%%%
%%%  Fig 12  %%%
%%%%%%%%%%%%%%%%
\begin{figure}
\centering
\includegraphics[width=\linewidth]{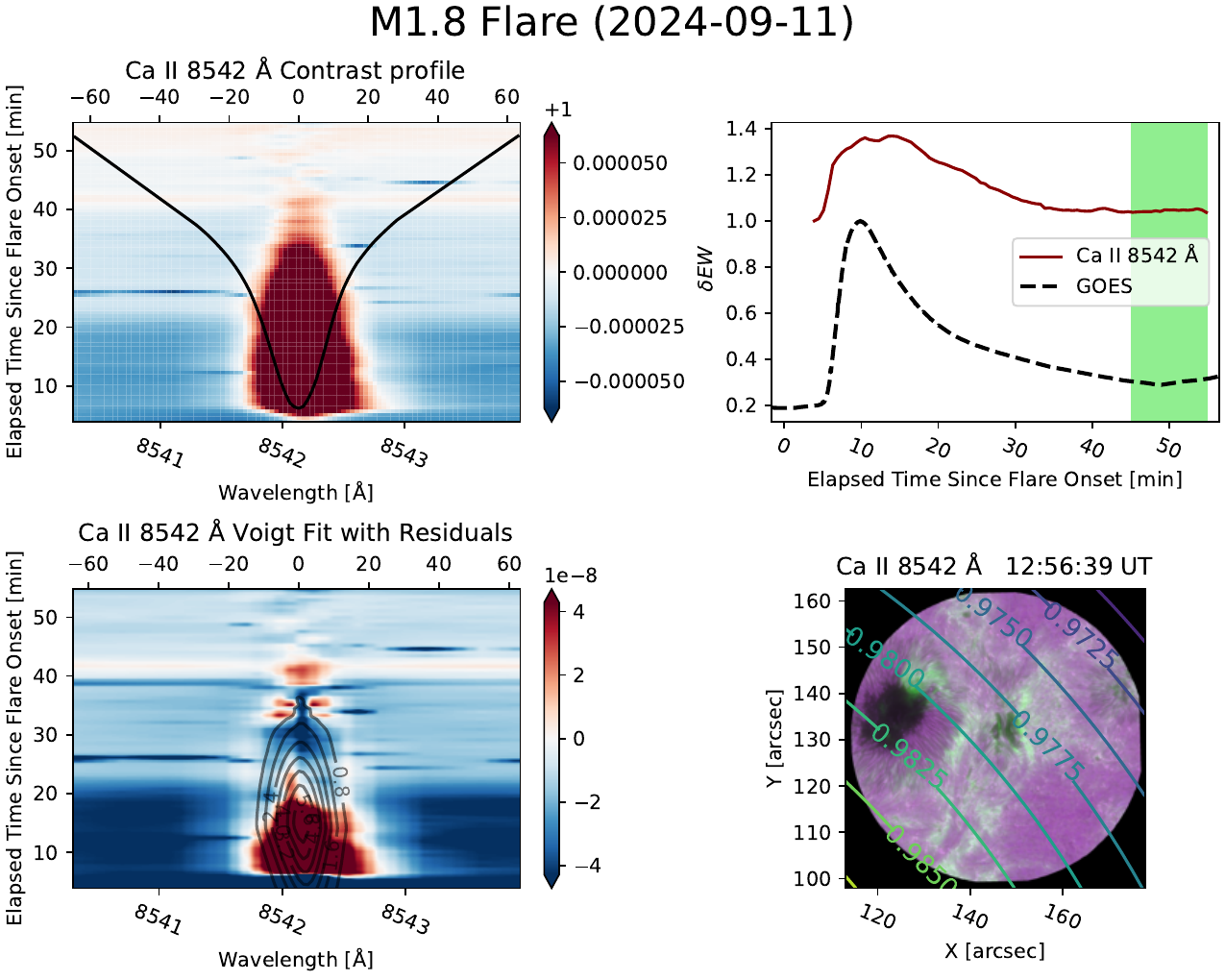}
\caption{Event 9. Same as Fig.~\ref{fig:X9.3} but for \CaIR.}
\label{fig:M1.8}
\end{figure}
%%%%%%%%%%%%%%%%%

%%%%%%%%%%%%%%%%
%%%  Fig 13  %%%
%%%%%%%%%%%%%%%%
\begin{figure*}
\centering
\includegraphics[width=0.75\textwidth]{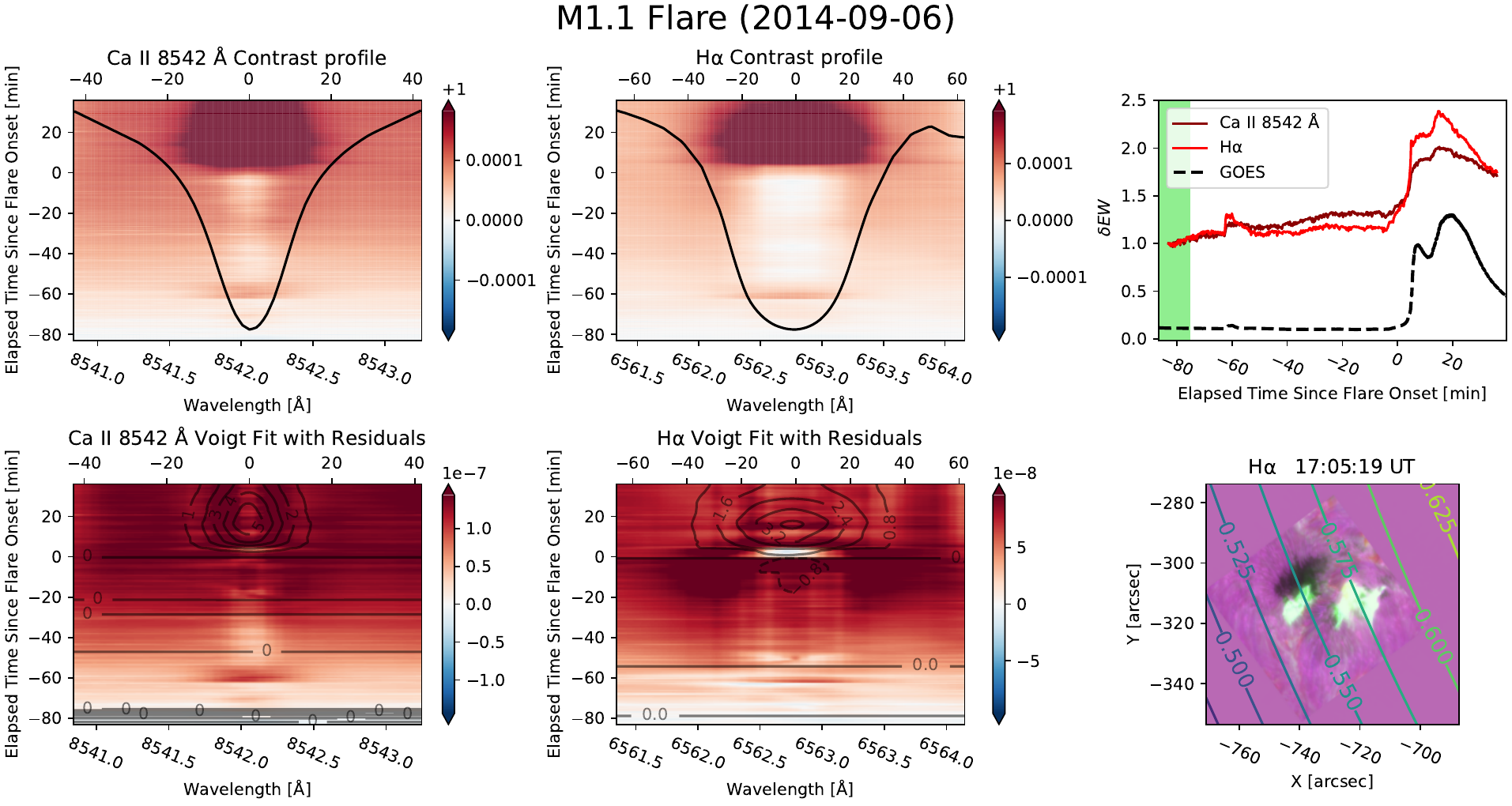}
\caption{Event~10. Same as Fig.~\ref{fig:X9.3} but for \CaIR\ and \Halpha.}
\label{fig:M1.1}
\end{figure*}
%%%%%%%%%%%%%%%%%

%%%%%%%%%%%%%%%%
%%%  Fig 14  %%%
%%%%%%%%%%%%%%%%
\begin{figure*}
\centering
\includegraphics[width=0.75\textwidth]{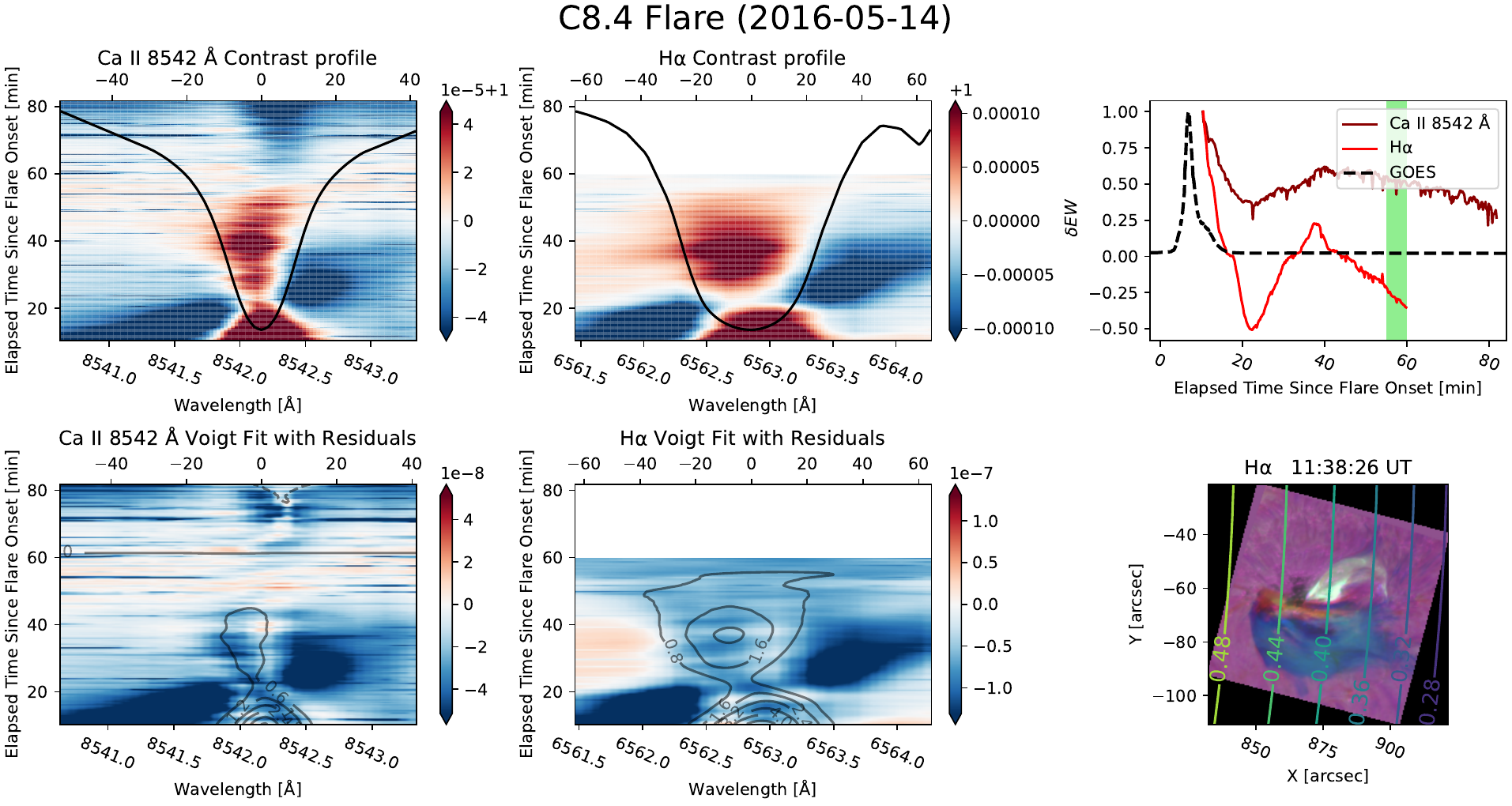}
\caption{Event~11. Same as Fig.~\ref{fig:X9.3} but for \CaIR\ and \Halpha. }
\label{fig:C8.4}
\end{figure*}
%%%%%%%%%%%%%%%

\subsubsection*{Event 9: An M1.8 flare on 2024 September 11}

An M1.8 flare occurred in the active region NOAA 13814 on 2024-09-11 at $\sim$12:27~UT. Approximately 80\% of the flare ribbon area was captured in the FOV of the SST. The average $\mu$ value of the flare is 0.979.  The event has an associated CME with 400~\kms. The analysis is displayed in Fig.~\ref{fig:M1.8}. 

This flare occurred in a very flare-productive region where it was the first of four M-class flares that day. Flare emission started at several locations almost simultaneously, shortly after a filament eruption that caused the associated CME. The flare consists of two main flare ribbons that merge with smaller flaring ribbons around them, much like Event~5. The gradual phase is relatively short and thus the ribbon emission dies down relatively quickly. The repeated subsequent flares in the same location indicate that the reconnection in this flare only partially relaxed the surrounding magnetic topology, which could explain the relatively brief gradual phase of this flare. 

This is the closest flare to the disk center. The filament and CME arose from material outside of the FOV. From the ribbons themselves we do not see much blue or redshifted material from this flare, nor do we see any up- or down-flowing hot or cool overlying material in the videos, which also suggests that this is a rather confined event. Therefore, we highlight that the chromospheric lines do not evidence the associated CME via evaporation or condensation signatures.

\subsubsection*{Event 10: An M1.1 flare on 2014 September 6}

An M1.1 flare occurred in the active region NOAA 12157 on 2014-09-06 at $\sim$17:00~UT. Approximately 100\% of the flare ribbon area was captured in the FOV of the SST. The average $\mu$ value of the flare is 0.56.  No CME was associated with this event. This flare was first described in \citet{Kuridze2015} and was observed during the IRIS and SST coordinated campaigns described in \citet{Luc2020}. 

This flare shows some striking differences to the M1.9 flare (Event~8). Both have similar strengths in GOES and occur near the limb. However, the emission from this flare is exceptionally symmetric in both the contrast profiles and the Voigt profile residuals (see Fig.~\ref{fig:M1.1}). In this instance, no overlying absorbing material is present to provide absorption with a significant net Doppler shift, which results in the highly symmetric profiles. This emphasizes the power of coronal rain, filaments, and fibrils in determining the net Doppler velocities of \mbox{Sun-as-a-star} flare observations of chromospheric lines, especially when contrasted with the flare ribbons themselves. The same misinterpretation can also occur in spatially resolved studies, where individual wavelength images are inspected without appropriate context, showing what emission originates from flare ribbon and what spectral features are linked to overlying material. Context can easily be provided by a COCOPLOT.

Exceptionally, the observation includes a rather long segment of preflare data, which includes a detected microflare at $-60$ minutes. We also observe a 50-minute line center darkening preceding the flare which then turns into faint red- and blue-shifted emission. The other investigated flares do not include long enough preflare data to check whether this is a common emission pattern, more pre-flare time series are needed to answer this question. 

%%%%%%%%%%%%%%%%
%%%  Fig 15  %%%
%%%%%%%%%%%%%%%%
\begin{figure*}
\centering
\includegraphics[width=0.75\textwidth]{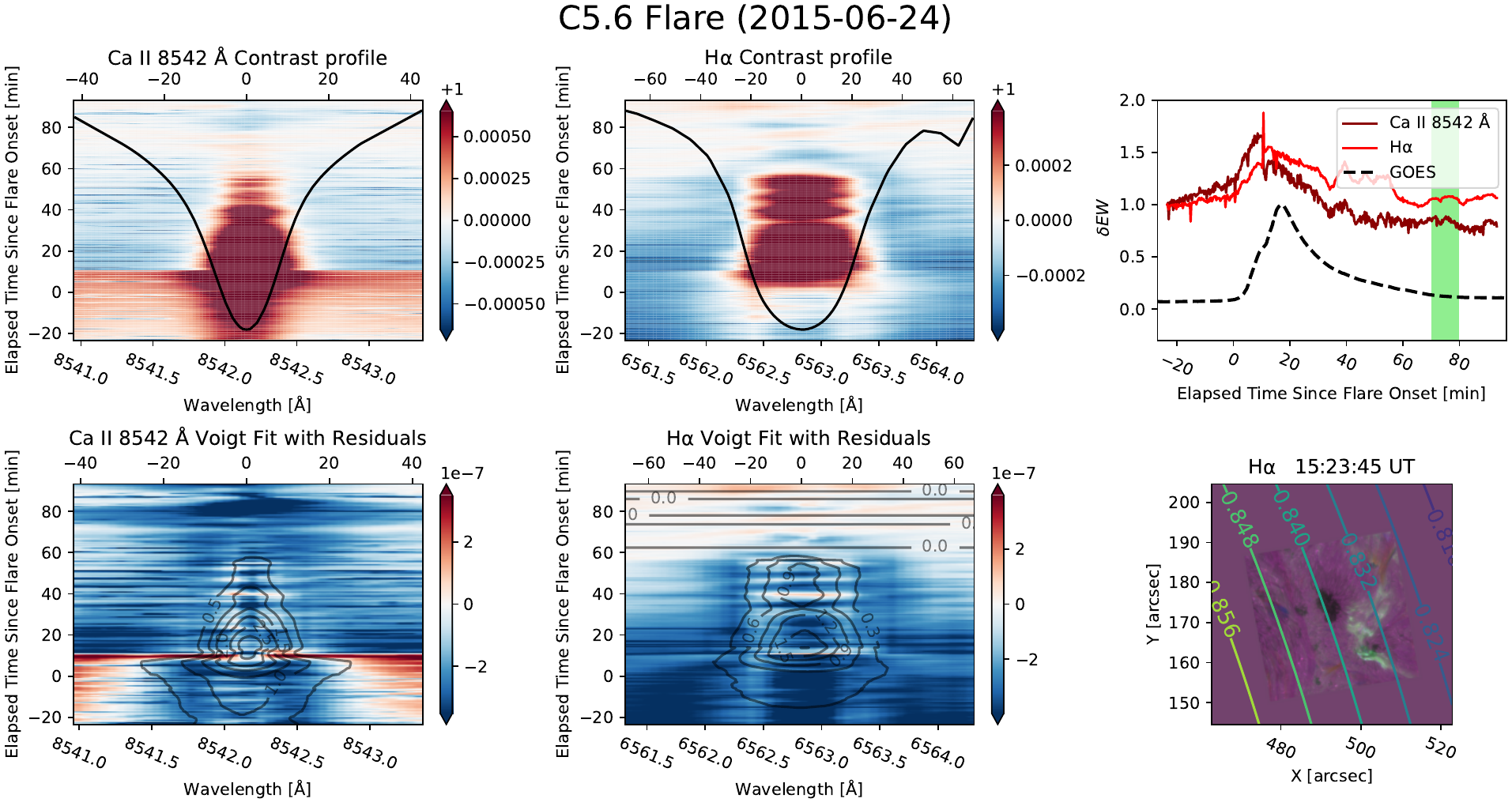}
\caption{Event~12. Same as Fig.~\ref{fig:X9.3} but for \CaIR\ and \Halpha.}
\label{fig:C5.6}
\end{figure*}
%%%%%%%%%%%%%%%%%

%%%%%%%%%%%%%%%%
%%%  Fig 16  %%%
%%%%%%%%%%%%%%%%
\begin{figure}
\centering
\includegraphics[width=\linewidth]{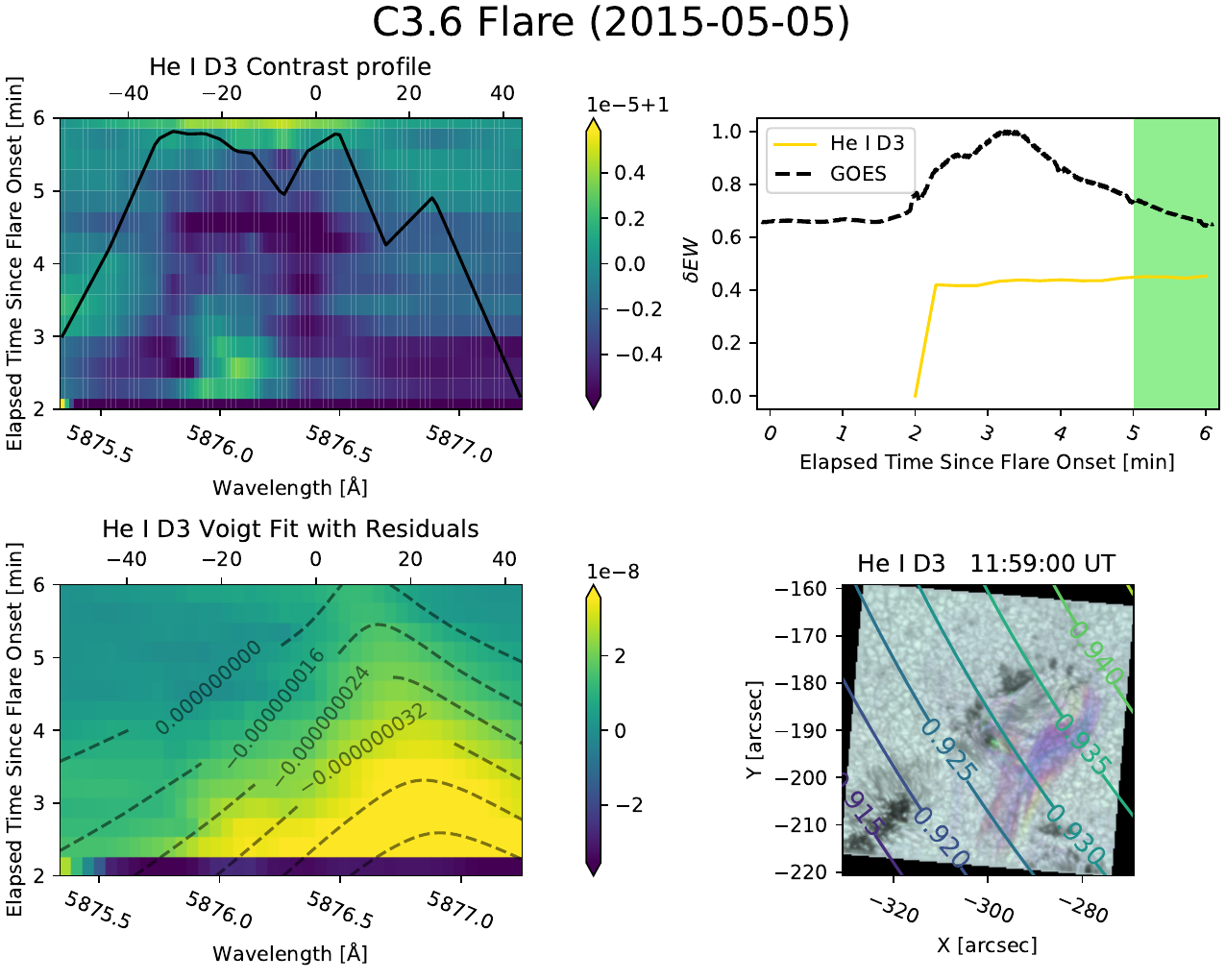}
\caption{Event~14. Same as Fig.~\ref{fig:X9.3} but for \HeD. In the top left panel, the absorption features in the \mbox{quiet-Sun} profile (black curve) are blends of telluric H$_2$O and solar metallic lines (see Fig.~2 in \citet{Libbrechtetal2017} and the top panels of Fig.~4.4 in \citet{Libbrecht2019}).The $\delta EW$ of \HeD\ has been offset by 1.}
\label{fig:C3.6}
\end{figure}
%%%%%%%%%%%%%%%%%

\subsubsection*{Event 11: A C8.4 flare on 2016 May 14}

A C8.4 flare occurred in the active region NOAA 12543 on \mbox{2016-05-14} at $\sim$11:28~UT at $\mu = 0.38$. The entirety (100\%) of the ribbon area was captured in the FOV of the SST. A small 144~\kms\ CME was associated with this event. This flare was first described in \citet{Kuridze2017}.

The impulsive phase of the flare started 5 to 10 minutes before the start of the SST observation (Fig.~\ref{fig:C8.4}). A small filament/prominence was present over the classic sigmoid \footnote{sigmoid flare ribbons are S-shaped, not $\sigma$ shaped.} flaring area. The cool prominence material produces strong absorption in chromospheric lines where it covers the flaring area in the line of sight at times near the start of the SST observation. This is especially evident in \Halpha\ observations where the prominence material is highly opaque (Fig.~\ref{fig:C8.4}, bottom right panel, strong blue and red features). In the first ten minutes of the SST observation (flare time $10-20$ minutes) the bring flare ribbon emission decreases, but there is also an evolution of the filament with \mbox{up-and-downflows} as well as the launch of some fraction as a CME. This reorganization and draining of the filaments clears the view of the flaring area leading to a clear second, more gradual "peak" in the flare emission (Fig.~\ref{fig:C8.4}, top right panel) which is particularly strong in \Halpha\ compared with \CaIR, again due to the high opacity of the filament in \Halpha.

The contrast profiles and Voigt residuals clearly show the effect of gravity acting on the overlying absorbing material in this filament from minutes $10-35$ from the deceleration of upflowing material (blue wing absorption of the spectral lines in Fig.~\ref{fig:C8.4}, left and central panels), and the acceleration of downflowing material (subsequent red wing absorption), which manifests in an approximately linear drift of the Doppler velocities to longer wavelength over time. \mbox{Back-of-the-envelope} calculations of the linear drift in Doppler velocity ($-60$ to 60~\kms\ over a time of 25 minutes, viewed at angle $\mu = 0.38$ give a constant acceleration of 0.210~km\,s$^{-2}$. Given the errors in this calculation and the angle of the magnetic field with respect to the vertical, this action is consistent with draining under solar gravity (0.274~km\,s$^{-2}$).  
There is a blue-shifted absorption feature seen in the \CaIR\ line core at times after $t\sim50$ minutes.  As in Event~7, this is a result of taking the reference frames for the contrast profile during the gradual phase of the flare.

%%%%%%%%%%%%%%%%
%%%  Fig 17  %%%
%%%%%%%%%%%%%%%%
\begin{figure*}
\centering
\includegraphics[width=0.75\textwidth]{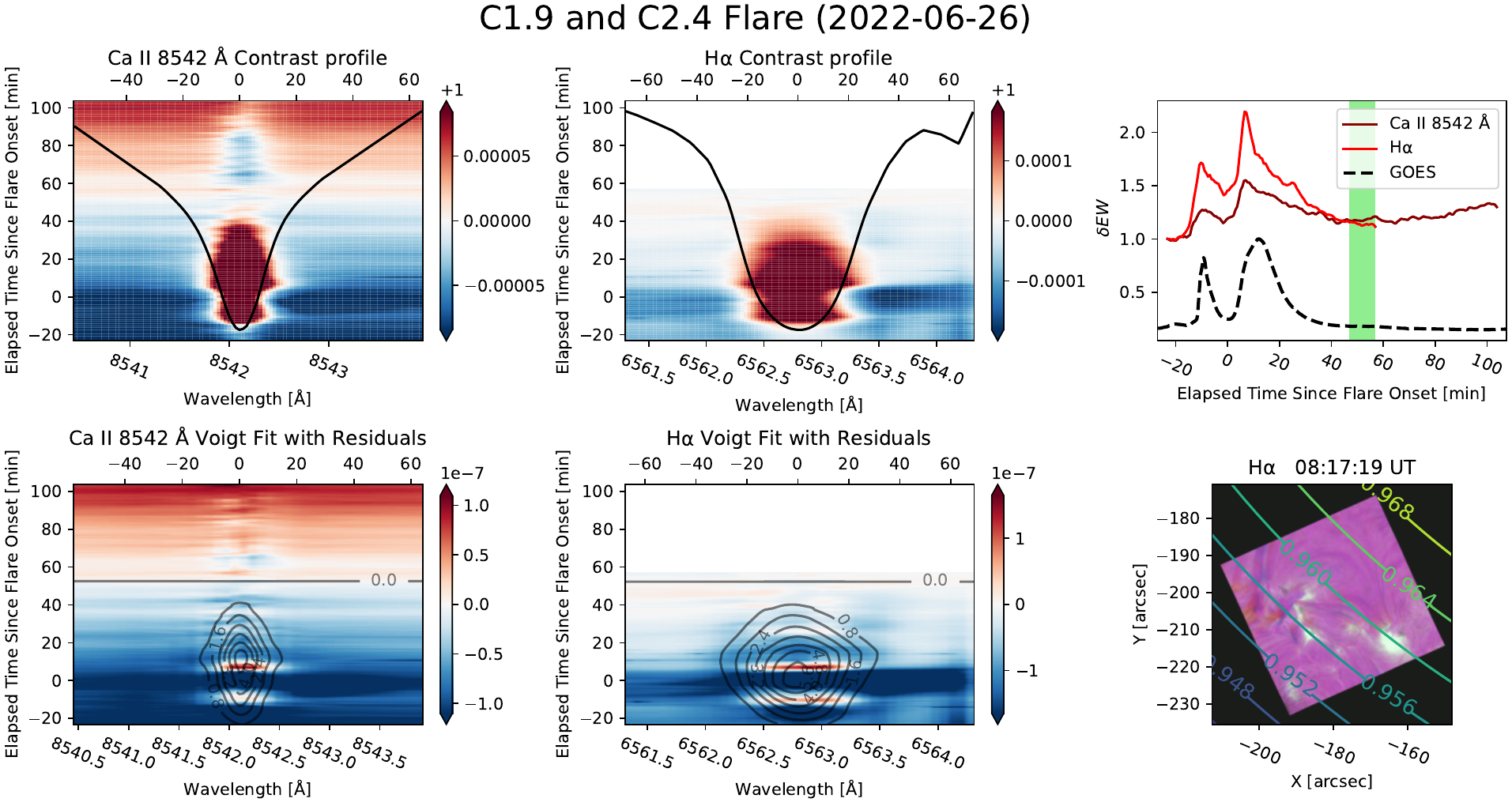}
\caption{Event~15. Same as Fig.~\ref{fig:X9.3} but for \CaIR\ and \Halpha.}
\label{fig:C1.9_and_C2.4}
\end{figure*}
%%%%%%%%%%%%%%%%%

%%%%%%%%%%%%%%%%
%%%  Fig 18  %%%
%%%%%%%%%%%%%%%%
\begin{figure*}
\centering
\includegraphics[width=\textwidth]{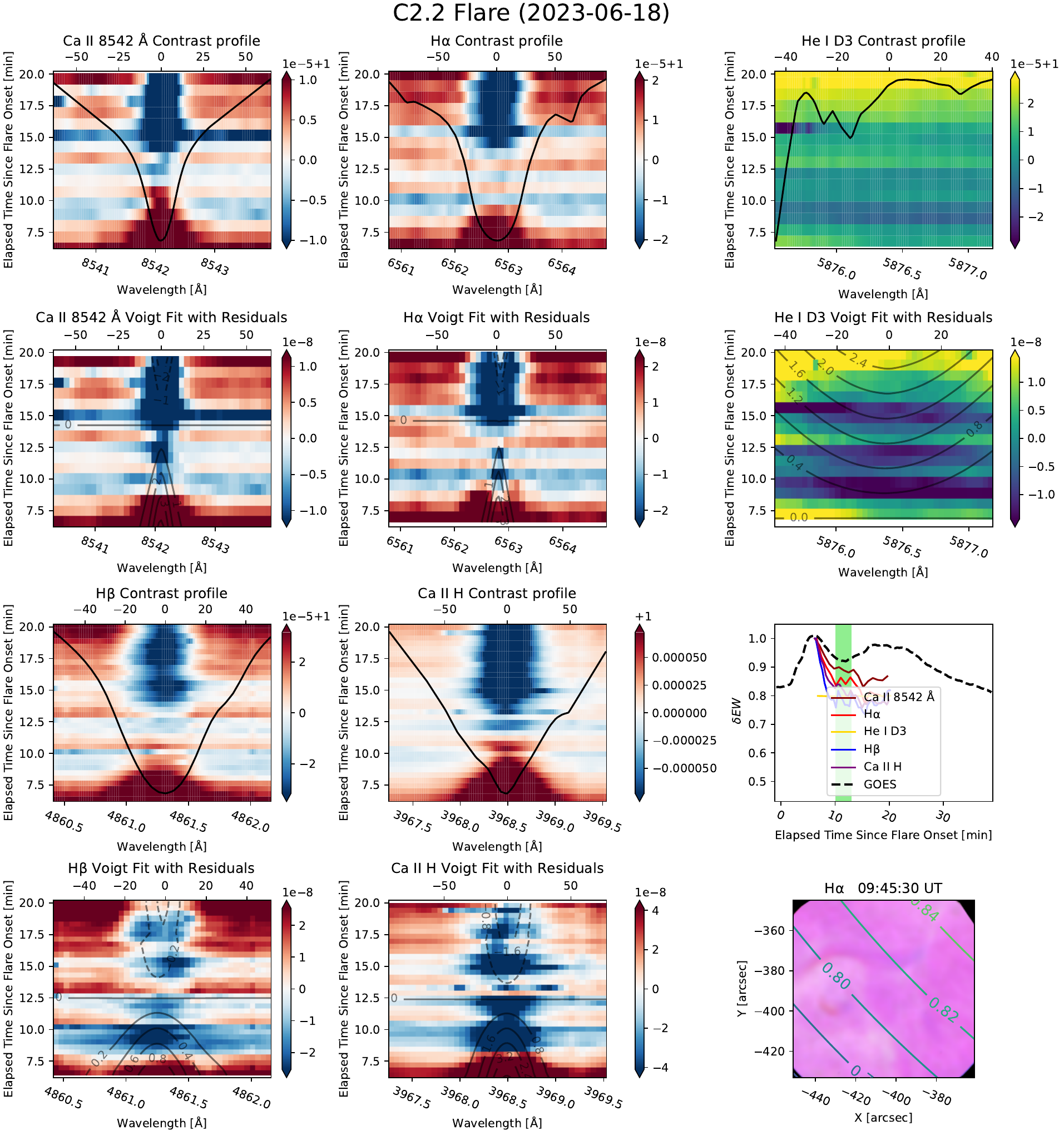}
\caption{Event~16. Same as Fig.~\ref{fig:X9.3} but for \CaIR, \Halpha, \HeD, \Hbeta, and \CaH. In the top right panel, the absorption features in the \mbox{quiet-Sun} profile (black curve) are blends of telluric H$_2$O and solar metallic lines (see Fig.~2 in \citet{Libbrechtetal2017} and the top panels of Fig.~4.4 in \citet{Libbrecht2019}).
Offset of $\delta EW$ of \HeD\ by 1.8.}
        \label{fig:C2.2}
    \end{figure*}
%%%%%%%%%%%%%%%%%

%%%%%%%%%%%%%%%%
%%%  Fig 19  %%%
%%%%%%%%%%%%%%%%
\begin{figure*}
\centering
\includegraphics[width=\textwidth]{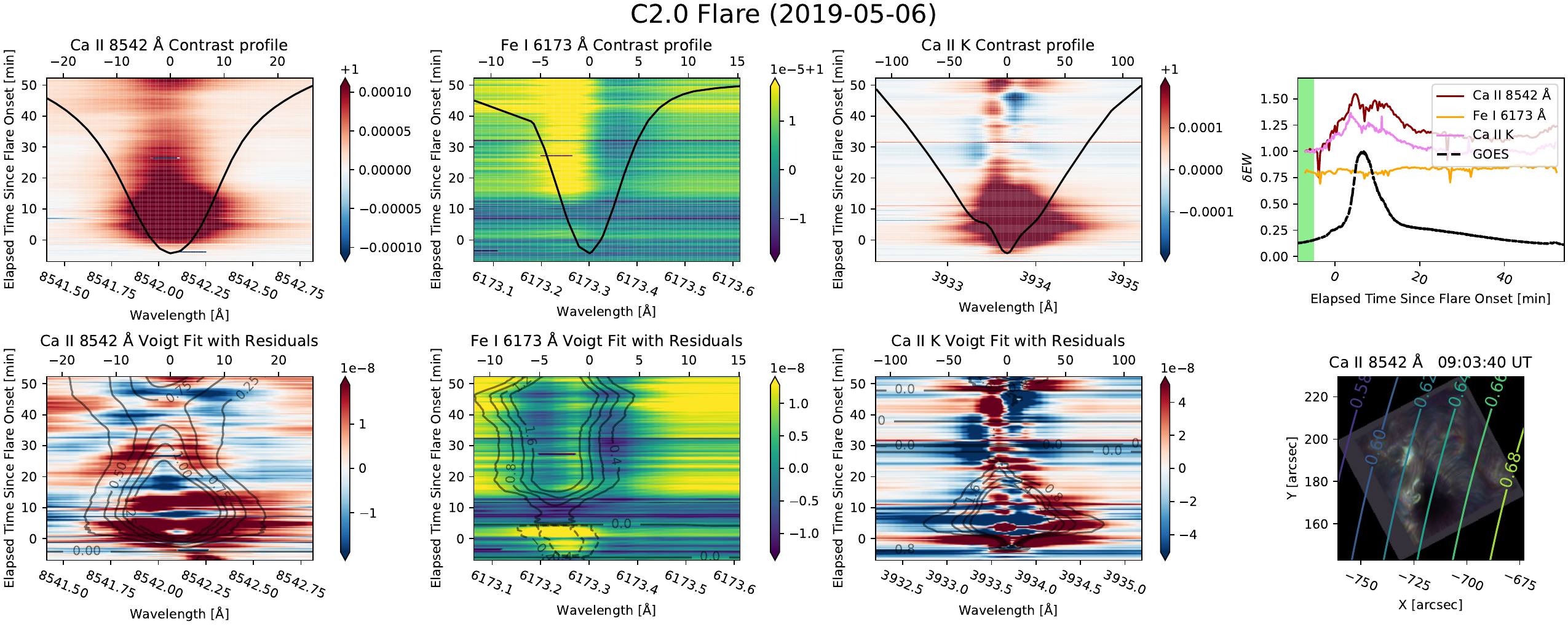}
\caption{Event~17. Same as Fig.~\ref{fig:X9.3} but for \CaIR, \Fe{i 6173}, and \CaK. Offset of $\delta EW$ of \Fe{i 6173} by 1.8.}
\label{fig:C2.0}
\end{figure*}
%%%%%%%%%%%%%%%%%

\subsubsection*{Event 12: A C5.6 flare on 2015 June 24}

An C5.6 flare occurred in the active region NOAA 12371 on 2015-06-24 at $\sim$15:30~UT. Approximately 10\% of the flare ribbon area was captured in the FOV of the SST from 5 minutes past the onset. The average $\mu$ value of the flare is 0.84. No CME is associated with this event. The analysis is displayed in Fig.~\ref{fig:C5.6}. 

This dataset changes its pointing around 30 minutes into the observation to better cover the flare. This causes a seeming discontinuity around the 10-minute mark. Hence, the normalization is taken after the readjustment. Several small jets and coronal rain can be seen throughout the FOV which manifest as red- and blue-shifted absorption around 60 minutes. The change in pointing at the time shortly before the GOES peak makes it difficult to strongly infer any delay in the timings of the peaks in emission of \Halpha, \CaIR, and GOES. 

\subsubsection*{Event 13: A C4.1 flare on 2011 August 6}

A C4.1 flare occurred in the active region NOAA 11267 on \mbox{2011-08-06} at $\sim$08:40~UT. Approximately 50\% of the flare ribbon area was captured in the FOV of the SST. The average $\mu$ value of the flare is 0.85. However, the only captured lines are \Fe{i 6302} and \Fe{i 5576} which are insensitive to flaring activity of this caliber. Therefore, this event was not pursued further. A detailed description of this event can be found in \citet{Cristaldi2014}.

\subsubsection*{Event 14: A C3.6 flare on 2015 May 5}

A C3.6 flare occurred in the active region NOAA 12335 on \mbox{2015-05-05} at $\sim$11:55~UT \citep{2019Libbrecht}. Approximately 95\% of the flare ribbon area was captured in the FOV of the SST, which observed only the \HeD\ line. The average $\mu$ value of the flare is 0.929. No CME is associated with this event. This flare was first described in \citet{2019Libbrecht}. The analysis is displayed in Fig.~\ref{fig:C3.6}. 

The observation starts with one bad frame as the telescope is still moving to capture the event (see Fig.~\ref{fig:C3.6}). 

In the contrast profile we can see some faint blueshifted emission, with a redshifted absorption next to it. The Voigt fit comes out negative due to a lack of emission core, and shows what seems to be a redshifted emission in the residuals, but this should in fact be interpreted as an absorption that slows down over time. Its presence aligns with the appearance and density of the flare feature seen in the data. 

\subsubsection*{Event 15: A double C1.9 and C2.4 flare on 2022 June 26}

A C1.9 flare followed by a C2.4 flare occurred in the active region NOAA 13040 on 2022-06-26 at $\sim$08:12~UT. Approximately 90\% of the second flare ribbon area was captured in the FOV of the SST. The average $\mu$ value of the flare is 0.959. No CME was associated with this event. A detailed description of this event can be found in \citet{Thoen2025}. The analysis is displayed in Fig.~\ref{fig:C1.9_and_C2.4}. 

The first flare happens partially out of the field of view and produces associated cool coronal loop material that does not become a CME. Instead the material falls back down, obscuring some part of the flare ribbons and creating a dual blue- and red-shifted absorption signatures. The second stronger flare starts toward the end of this process. From the pattern of the EWs alone it is not possible to tell whether this is one larger flare with emission that gets strongly absorbed near its start, or two smaller flares happening at relatively close times. In a true disk-integrated setting such an event could be mistaken for a single stronger flare if no X-ray or white light information is available. In \CaIR we see dimming in the contrast profile core wavelengths after the flare in a fashion similar to Event~7. 

\subsubsection*{Event 16: A C2.2 flare on 2023 June 18}

A C2.2 flare occurred in the active region NOAA 13336 on \mbox{2023-06-18} at $\sim$09:31~UT. Approximately 100\% of the flare ribbon area was captured in the FOV of the SST. The average $\mu$ value of the flare is 0.808. No CME was associated with this event. The analysis is displayed in Fig.~\ref{fig:C2.2}. 

This flare was once again captured several minutes after its onset, and thus the SST observations begin near the event peak. Note that the contrast is taken relative to a time during the flare, this confuses interpreting the contrast profiles and Voigt fit residuals. All lines initially show a very similar pattern with a central brightening. After minute 10, a naive interpretation would be of a central dimming, which starts at minute 14 and continues throughout the rest of the time series, which is comparatively very short. However, in reality this represents a gradual dimming of the core enhancement which has its peak near the start of the flare. The horizontal "bands" visible in the contrast profiles and Voigt fits are due to variable seeing conditions.

We see dimming in the contrast profile core wavelengths later in the flare in a fashion similar to Event~7 due to taking the reference profile at 10-15 minutes. 

\subsubsection*{Event 17: A C2.0 flare on 2019 May 6}

A C2.0 flare occurred in the active region NOAA 12740 on \mbox{2019-05-06} at $\sim$08:45~UT. Approximately 100\% of the flare ribbon area was captured in the FOV of the SST. However, this alters through time and there is substantial in and outflow in the FOV by the rotational nature of the flare. The average $\mu$ value of the flare is 0.64. A weak 128~\kms\ CME is associated with this event. The flare was first described in \citet{Yadav2021}. The analysis is displayed in Fig.~\ref{fig:C2.0}. 

To better capture the event, the FOV was moved two times during this time series, causing jumps in the contrast profiles. These jumps took place around 5 and 32 minutes and caused some visible distortions in the profiles. Besides this, we can see the characteristic red-shifted emission caused by the flare ribbon formation processes compacting pre-existing chromosphere structures (as in Event~7). Any effects due to blue os redshifted cool overlying material later on in the time series are difficult to discern due to the impact on the net emission of the shift in telescope pointing. 

\subsubsection*{Event 18: A C1.5 flare on 2013 June 30}
    
\begin{figure}
\centering
\includegraphics[width=\linewidth]{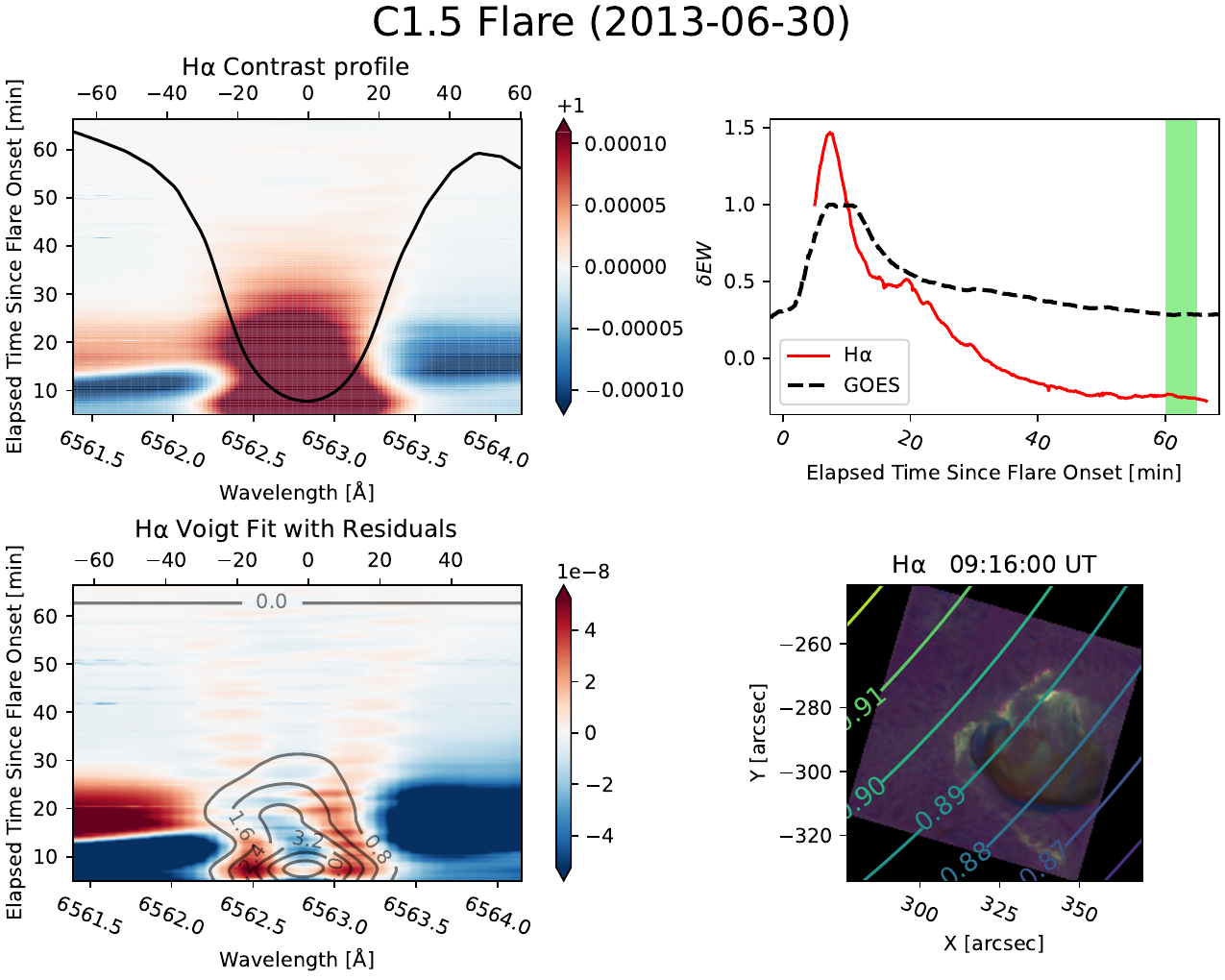}
\caption{Event~18. Same as Fig.~\ref{fig:X9.3} but for \Halpha.}
\label{fig:C1.5}
\end{figure}

A C1.5 flare occurred in the active region NOAA 11778 on \mbox{2013-06-30} at $\sim$09:20~UT, associated with a confined filament eruption \citep{2019DoyleC1.5} but no CME. Full capture of the flare ribbon area in the FOV of the SST is assumed. The average $\mu$ value of the flare is 0.64. This flare was first described in \citet{2017Druett}. The analysis is displayed in Fig.~\ref{fig:C1.5}. 

This flare is covered by the aforementioned filament eruption which launches a large amount of material up toward the observer, leaving a strong blue-shifted excess absorption signature near the start of the observation. Around 15 minutes into the event, a significant fraction of this overlying material falls back down as coronal rain, which is clearly seen via absorption in the red wing of the contrast profiles. The flare ribbon emission below becomes more visible with time due to the decreasing overlap with overlying material. This reveals some red-shifted emission due to flare ribbon formation as in Event~7. Each of these physical processes is clearly and simply manifested in the contrast profiles, partially aided by a long observation sequence that provides a suitable stable reference profile to help construct easily interpretable contrast profiles. However, the previously described flare events illustrate that such a clear correspondence between spectral signature and physical motions does not always hold true. 

The oscillations in intensity are once again seeing induced. 

%%%%%%%%%%%%%%%%
%%%  Fig 20  %%%
%%%%%%%%%%%%%%%%
%%%%%%%%%%%%%%%%

\subsubsection*{Event 19: A C1.2 flare on 2021 May 26}
    
\begin{figure*}
\centering
\includegraphics[width=\textwidth]{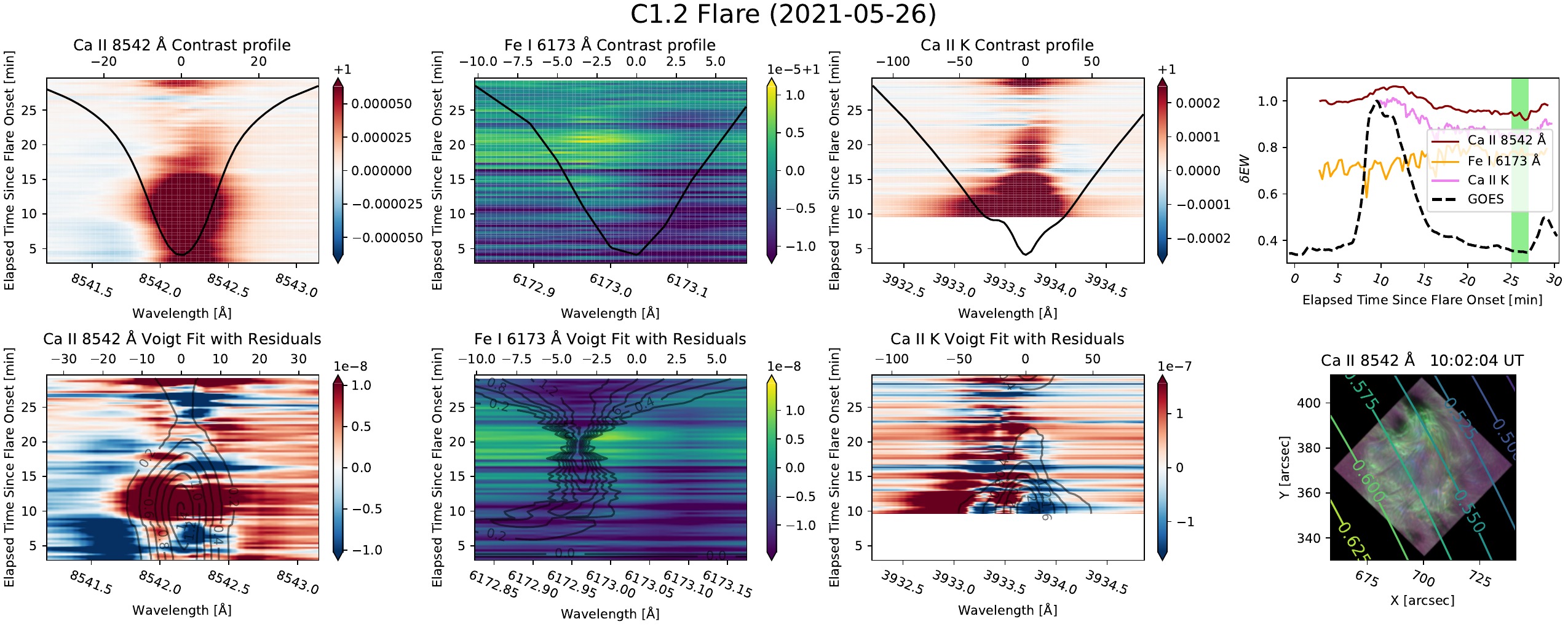}
\caption{Event~19. Same as Fig.~\ref{fig:X9.3} but for \CaIR, \Fe{i 6173}, and \CaK. Offset of $\delta EW$ of \Fe{i 6173} by 1.7.}
\label{fig:C1.2}
\end{figure*}

A C1.2 flare occurred in the active region NOAA 12826 on \mbox{2021-05-26} at $\sim$09:55~UT. Approximately 100\% of the flare ribbon area was captured in the FOV of the SST. The average $\mu$ value of the flare is 0.56. No CME is associated with the event. The analysis is shown in Fig.~\ref{fig:C1.2}. 

This event represents a particularly weak flare inside of a decaying active region that produced several more C-class flares in the following days before it rotated out of sight behind the limb. In this case, one primarily sees mildly tangled field lines linking the fragmented spots and pores of the decaying region. The flare occurs at the light bridge in the upper corner of the observation and starts off very delicately with several small \mbox{micro-flare} brightenings before a larger reconnection occurs around 7 minutes. Around 15 minutes the reconnection calms down before several new microflares occur around 20 minutes and toward the end of the time series.

The seeming blueshift visible in the iron line seems to be primarily caused by changes in the FOV pointing during the time series, as it slowly rotates in a way that increases the amount of penumbra on one side of the FOV, while cutting it off on the other. %wait for movie to check 22min

%%%%%%%%%%%%%%%%
%%%  Fig 21  %%%
%%%%%%%%%%%%%%%%
%%%%%%%%%%%%%%%%

%%%%%%%%%%%%%%%%
%%%  Fig 22  %%%
%%%%%%%%%%%%%%%%
\begin{figure}
    \centering
    \includegraphics[width=\linewidth]{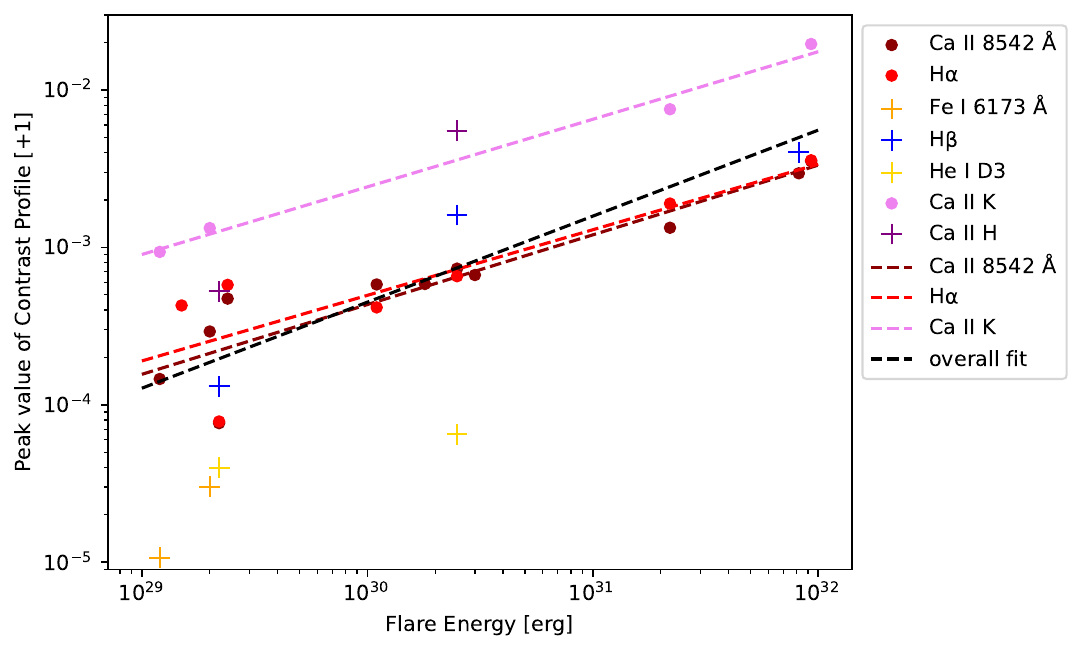}
    \caption{Correlation between peak values of contrast profiles and flare energy. Lines with four or more data points were fitted individually, while lines with fewer were only considered for the overall fit.
    Lines with individual fits are \CaIR, \Halpha, \CaK\ lines (dots) and lines only considered for the overall fit are \Fe{i 6173}, \HeD, \Hbeta, \CaH\ lines (plus signs).}
    \label{fig:scalelaw}
\end{figure}
%%%%%%%%%%%%%%%%

\subsection{Flare energy scaling law}

The quantitative results presented in the previous section motivated us to explore a possible correlation between the peak values of contrast profiles and the radiated bolometric flare energy (hereafter the flare energy). We assumed a linear relationship between energy and soft \mbox{X-ray} flux in the $1-8$\,\AA\ passband, as measured by GOES, setting an X1.0 flare to $10^{31}$\,erg (e.g., Fig.~2 in \citet{Shibata2013} and Fig.~5 in \citet{Maehara2015}). Fig.~\ref{fig:scalelaw} illustrates the correlation between the peak values of contrast profiles, $C_\mathrm{peak}$, and flare energy, $E$. The $C_\mathrm{peak}$ values for spectral lines with four or more measurements (i.e., \CaIR, \Halpha, and \CaK) are represented by dots, while those with fewer than four measurements (\Fe{i 6173}, \HeD, \Hbeta, and \CaH) are shown as plus signs. Events~4, 11, and 14 are excluded as we cannot confirm that the observations captured the peak in $\delta EW$. 
Events~5, 8, and 12 exhibit less than 30\% overlap between the FOV and the flare and are therefore excluded, as scaling the intensity by the inverse of the overlap ($O^{-1}$ in Eq.~\ref{eq:delta_I}) introduces large errors. Fig.~\ref{fig:scalelaw} suggests a \mbox{power-law} relationship of the form \mbox{$C_\mathrm{peak} \sim E^a$}, which can be expressed as \mbox{$\log_{10}C_\mathrm{peak} \sim a\log_{10}E$}. The corresponding linear fits for the values $C_\mathrm{peak}$ with four or more measurements are depicted by dashed lines, with colors that match the respective spectral lines. The overall fit, which incorporates all values of $C_\mathrm{peak}$, is shown as a black dashed line. Correlation coefficients and fitting parameters are listed in Table~\ref{tab:summary_fit}. The slope values $a$ suggest that the peak values of contrast profiles approximately scale as the square root of flare energy. 

%%%%%%%%%%%%%%%%%%%%%%%%%%%%
%%%      Table 2         %%%
%%%%%%%%%%%%%%%%%%%%%%%%%%%%
\begin{table}%[htbp]
    \centering
    \caption{Correlation coefficients ($R$) and fitting parameters for the function $\log_{10}(C_{\text{peak}}) = b + a \log_{10}(E)$ corresponding to the data in Fig.~\ref{fig:scalelaw}.}
    \label{tab:summary_fit}
    \begin{tabular}{lccc}
        \toprule
        \toprule
        Line & $R$  &  $b$  & $a$ \\
        \midrule
        \CaK        & 0.995 & $-15.5$ & 0.430 \\
        \CaIR       & 0.924 & $-16.6$ & 0.443 \\
        \Halpha     & 0.847 & $-15.8$ & 0.417 \\        
        \midrule
        Overall Fit & 0.744 & $-19.7$ & 0.547 \\
        \bottomrule
    \end{tabular}
\end{table}
%%%%%%%%%%%%%%%%%%%%%%%%%%%%

\section{Conclusions}

In this work, we have  used a unique collection of 20 solar flares observed with the SST (see Fig.~\ref{fig:sun_with_observations}) to obtain statistically relevant results. These limited-FOV time series were averaged and turned into approximations of \mbox{disk-integrated} spectra using the NESSI code \citep{2023PietrowNESSI} which were shown to closely reproduce true \mbox{Sun-as-a-star} observations with certain caveats such as the incomplete capture of a flare, seeing variations, as well as FOV shifts, jumps and drifts. As noted, limited-FOV observations are prone to errors introduced by quiet-Sun areas adjoining sunspots or active regions. Our method mitigates these issues while exploiting the main advantage of a restricted FOV--namely, isolating the spectral influence of the flaring region itself. Partially covered flares therefore remain informative, because the behavior they reveal still contributes to the flare’s overall spectral evolution; however, caution is warranted, as unobserved portions of the event may introduce unaccounted-for effects.

While only the strongest events from our sample have enough S/N to show up in true \mbox{Sun-as-a-star} and stellar observations, we believe that there is still merit in observing weaker events as their higher occurrence rate allows for better statistics, especially since these are physically the same events. 

Overall we note that the imprint of flares on the \mbox{Sun-as-a-star} spectrum is strongest in the \CaHK\ lines, which is in line with the findings of \citet{2024pietrowHarps}. This means that these lines are crucial for identifying activity that may not be as clearly present in other lines. 

% Feature summary
The analysis of contrast and residual profiles across a broad sample of flares reveals several recurring spectral features. The most prevalent is core intensity enhancement (Feature~1), or absorption in the case of \HeD, associated with the flare peak. Its presence is typically accompanied by line broadening and an enhancement in the normalized differenced equivalent width ($\delta EW$). A second frequent feature is transient redshifted emission (Feature~2), attributed to down-flowing plasma, likely coronal rain, in flare loops, most clearly detected at viewing angles that maximize the line-of-sight velocity component. Filament and flare loop imprints (Feature~3) manifest in a subset of events as dynamic reversals in wing asymmetry, driven by evolving structures obscuring the flare ribbons. Finally, several events show a temporal offset between $\delta EW$ and GOES peak times (Feature~4), suggesting variability in the timing of chromospheric response relative to coronal emission; however, incomplete temporal coverage in some cases limits definitive conclusions.

% CMEs
Of the sampled flares, $53$\% had cataloged CMEs, yet only one (Event~11) showed distinct spectral signatures resembling known CME-associated profiles—highlighting the lack of a consistent spectral fingerprint for CME detection. Several non-CME events exhibited CME-like Doppler features caused by gravitationally bound upflows, emphasizing the risk of false positives. These findings underscore the limitations of using flare presence or Doppler signatures alone to infer CMEs and call for further research to improve the reliability of stellar CME diagnostics. 

Pseudo Sun-as-a-star analysis of flares has so far mainly been focused on individual cases \citep{Namekata2022Saas, Xu2022, Otsu2024, Otsu2024b}, with the exception of \citet{Otsu_2022} who studied seven events, and \citet{Yu2025} who took a more general simulated approach. We believe that our work represents a large step in this field, due to the number of studied events that make it possible for statistically significant commonalities to be found, although we believe that even larger studies are necessary. Our simulated full-disk observations using the NESSI code also allowed us to have accurate intensities for the contrast profiles and residuals, which have often been lacking in previous works. 

Additionally, this work allowed us to validate these previous studies, as our flares are observed at a much higher spatial and temporal resolution. Perhaps unsurprisingly, we do not report a strong gain from the resolution increase, but we could temporally resolve much more of the flare structure. Additionally, the multi-line observations used in this work provided additional information that is often missing from single-line observations. 

We believe that the spectral features described in this paper can help interpret unresolved stellar flares, as well as show that strong blue-shifted absorption does not necessarily indicate a CME. However, we also urge for more studies of this type with larger amounts samples of flares to help solidity these findings. 

\subsection{Data availability}

The datasets in this work are collected from a large number of observers and institutes, with only some of them being public. Please contact the corresponding author for more information on specific sets. 

\begin{acknowledgements}
%Personal
This project is the result of an internship led by AP and MD.

%% -- AP
AP is supported by the \emph{Deut\-sche For\-schungs\-ge\-mein\-schaft, DFG\/} project number PI 2102/1-1

%% -- MD
MD is supported by FWO project G0B4521N.
MD acknowledges funding from the European Research Council (ERC) under the European Union Horizon 2020 research and innovation program (grant agreement No.~833251 PROMINENT ERC-ADG 2018). 
The computational resources and services used in this work were provided by the VSC (Flemish Supercomputer Center), funded by the Research Foundation Flanders (FWO) and the Flemish Government, department EWI.

%% -- APY
APY acknowledges support from the Swedish Research Council (grant 2023-03313). APY acknowledges funding by the European Union through the European Research Council (ERC) under the Horizon Europe program (MAGHEAT, grant agreement 101088184). 

%% -- DNS
DNS acknowledges support by the European Research Council through the Synergy Grant number 810218 ("The Whole Sun", ERC-2018-SyG).

%% -- JdlCR
JdlCR gratefully acknowledges funding by the European Union through the European Research Council (ERC) under the Horizon Europe program (MAGHEAT, grant agreement 101088184).  The Institute for Solar Physics is supported by a grant for research infrastructures of national importance from the Swedish Research Council (registration number 2021-00169). 

%% -- ARB, JTF, LRvdV, ESØ, and RJ
ARB, RJ, LRvdV, JTF, and ESØ are supported by the Research Council of Norway (RCN) through its Centres of Excellence scheme, project number 262622. LRvdV acknowledges support by RCN project number 325491.

%% -- AMS
AMS is funded by the European Union (ERC, FIERCE, 101052347). Views and opinions expressed are however those of the author(s) only and do not necessarily reflect those of the European Union or the European Research Council. Neither the European Union nor the granting authority can be held responsible for them. This work was supported by FCT - Funda\c{c}\~{a}o para a Ci\^{e}ncia e a Tecnologia through national funds by these grants: UIDB/04434/2020 DOI: 10.54499/UIDB/04434/2020, UIDP/04434/2020 DOI: 10.54499/UIDP/04434/2020.

%% -- JK, JR, AW
JK, JR, and AW acknowledge the project VEGA~2/0043/24.

%% -- AB
The work of AB was supported by the program "Excellence \mbox{Initiative-Research} University" for the years \mbox{2020–2026} at the University of Wroc\l{}aw, project No. BPIDUB.4610.96.2021.KG. and from the project RVO:67985815 of the Astronomical Institute of the Czech Academy of Sciences.

%% -- DK
DK acknowledges the Science and Technology Facilities Council (STFC) grant ST/ W000865/1 to Aberystwyth University and the Georgian Shota Rustaveli National Science Foundation project FR-22-7506.

%% -- ZV
ZV is supported by the Fellowship for Excellent Researchers \mbox{R2-R4} (\mbox{09I03-03-V04-00015}). 

%% -- TVZ
TVZ was supported by the Austrian Science Fund (FWF) project PAT7550024, by Shota Rustaveli National Science Foundation of Georgia (project FR-21-467) and by the International Space Science Institute (ISSI) in Bern, through the ISSI International Team project 24-629 (\mbox{Multi-scale} variability in solar and stellar magnetic cycles).

%% -- SST
The Swedish 1-m Solar Telescope is operated on the island of La Palma by the Institute for Solar Physics of Stockholm University in the Spanish Observatorio del Roque de los Muchachos of the Instituto de Astrof\'{\i}sica de Canarias. The Swedish 1-m Solar Telescope, SST, is co-funded by the Swedish Research Council as a national research infrastructure (registration number 4.3-2021-00169). 

%% -- SOLARNET
Some research data leading to the results obtained has been supported by the SOLARNET project that has received funding from the European Union’s Horizon 2020 research and innovation programme under grant agreement no 824135.

%% -- F-CHROMA
Some research data leading to the results obtained has been supported by the \mbox{F-CHROMA} database (\href{https://star.pst.qub.ac.uk/wiki/public/solarmodels/start.html}{https://star.pst.qub.ac.uk/wiki/public/solarmodels/start.html}), which was funded by the European Community’s Seventh Framework Programme (\mbox{FP7/2007-2013}) under grant agreement no. 606862 (\mbox{F-CHROMA}), and from the Research Council of Norway through the Programme for Supercomputing.

%CME catalog
This CME catalog is generated and maintained at the CDAW Data Center by NASA and The Catholic University of America in cooperation with the Naval Research Laboratory. SOHO is a project of international cooperation between ESA and NASA.

\end{acknowledgements}

\bibliographystyle{aa}
\bibliography{aa54870-25}

\begin{appendix}
\section{Event table}
    
%%%%%%%%%%%%%%%%%%
%%%   Table 1  %%%
%%%%%%%%%%%%%%%%%%
\begin{table*}[b]
    \renewcommand{\tabcolsep}{0.9mm} 
     \renewcommand{\arraystretch}{1.05}
    \tiny    
    \caption{Event List}
    \label{tab:flare_info} 
    \centering
    \begin{tabular}{c c c r r c c r r@{ }r l l r@{.}l}
    \toprule
    \toprule
    No. & Date\tablefootmark{a} & NOAA\tablefootmark{b} & Class\tablefootmark{b} & $V_{\rm CME}$\tablefootmark{c}  & Flare Start\,/\,Peak\,/\,Stop\tablefootmark{b} &  
    SST Start\,/\,Stop & Overlap\tablefootmark{d} & \multicolumn{2}{c}{$(X, Y)$\tablefootmark{e}} & $\mu$ & Lines & \multicolumn{2}{c}{Cad.} \\
        &      &                       &                        &   (\kms)                        &                  (UT)                    &        (UT)      &                          &  \multicolumn{2}{c}{(\arcsec)}                &       &       & \multicolumn{2}{c}{(s)} \\
    \midrule    
%%%  No 1
    & & & & & & &  & & & & \CaIR & 15 & 0 \\
    1 & 2017-09-06 &  12673 & X9.3 & 1571 & 11:53\,/\,12:02\,/\,12:10 & 11:56\,/\,12:52 & $\sim40$\% & (513, & $-216)$ & 0.815 & \Halpha & 15 & 0 \\     
    & & & & & & & & & & & \CaK & 6 & 6 \\%[1.5mm] 
    \cmidrule{12-14}
%%%  No 2
    \multirow{4}{*}{2} & \multirow{4}{*}{2017-09-10} & \multirow{4}{*}{12673} & \multirow{4}{*}{X8.2} & \multirow{4}{*}{3163} & \multirow{4}{*}{15:35\,/\,16:06\,/\,16:31} & 16:23\,/\,18:32 & \multirow{4}{*}{$\sim50$\%} & \multirow{4}{*}{(959,} & \multirow{4}{*}{$-144)$} & \multirow{4}{*}{0.124} & \CaIR & 46 & 1 \\
    & & & & & & 16:23\,/\,18:32 & & & & & \Fe{i 6302} & 46 & 1 \\
    & & & & & & 16:25\,/\,17:49 & & & & & \Hbeta      & 21 & 5 \\
    & & & & & & 16:25\,/\,17:49 & & & & & \CaHK       & 19 & 7 \\%[1.5mm]
    \cmidrule{12-14}  
%%%  No 3            
    & & & & & & & & & & & \CaIR & 15 & 0 \\
    3 & 2017-09-06 & 12673 & X2.2 & 391 & 08:57\,/\,09:10\,/\,09:17 & 09:04\,/\,09:54 & $\sim50$\% & (542, &  $-208)$ & 0.803 & \Halpha & 15 & 0 \\         
    & & & & & & & & & & & \CaK & 6 & 6 \\%[1.5mm] 
    \cmidrule{12-14}
%%%  No 4   
    4 & 2014-06-10 & 12087 & X1.5 & 925 & 12:36\,/\,12:52\,/\,13:03 & 12:40\,/\,12:58 & $\sim75$\% & $(-879,$ & $-305)$ & 0.139 & \Halpha & 4 & 0\\%[1.5mm]     
    \cmidrule{12-14}
%%%  No 5  
    \multirow{2}{*}{5\tablefootmark{\dag}} & \multirow{2}{*}{2021-10-28} & \multirow{2}{*}{12887} & \multirow{2}{*}{X1.0} & \multirow{2}{*}{1519} & \multirow{2}{*}{15:17\,/\,15:35\,/\,15:48} & \multirow{2}{*}{15:32\,/\,16:08} & \multirow{2}{*}{$\sim10$\%} & \multirow{2}{*}{(150,} & \multirow{2}{*}{$-550)$} & \multirow{2}{*}{0.804} & \CaIR & 21 & 6\\%[1.5mm] 
    & & & & & & & & & & & \Fe{i 6173} & 21 & 6\\%[1.5mm]     
    \cmidrule{12-14}
%%%  No 6
    6\tablefootmark{\dag} & 2022-05-20 & 13014 & M3.0 & -- & 07:35\,/\,07:45\,/\,07:49 & 07:45\,/\,08:41 & $\sim90$\% & $(-11,$ & 358) & 0.927 & \CaIR & 21 & 5\\%[1.5mm]     
    \cmidrule{12-14}
%%%  No 7        
    & & & & & & & & & & & \CaIR & 53 & 7 \\
    & & & & & & & & & & & \Halpha & 53 & 7 \\   
    7\tablefootmark{\dag} & 2023-06-09 & 13331 & M2.5 & 651 & 16:48\,/\,17:11\,/\,17:34 & 17:01\,/\,17:52 & $\sim40$\% & $(-598,$ & $-353)$ & 0.688 & \HeD & 53 & 7 \\    
    & & & & & & & & & & & \Hbeta & 25 & 0 \\  
    & & & & & & & & & & & \CaH & 25 & 0 \\%[1.5mm]        
    \cmidrule{12-14}    
%%%  No 8                    
   \multirow{2}{*}{8} & \multirow{2}{*}{2015-09-27} & \multirow{2}{*}{12423} & \multirow{2}{*}{M1.9} & \multirow{2}{*}{--} & \multirow{2}{*}{10:20\,/\,10:40\,/\,10:46} & \multirow{2}{*}{10:35\,/\,10:59} & \multirow{2}{*}{$\sim10$\%} & \multirow{2}{*}{(775,} & \multirow{2}{*}{$-233)$} & \multirow{2}{*}{0.502} & \CaIR & 32 & 3 \\ 
   & & & & & & & & & & & \Halpha & 32 & 3 \\%[1.5mm] 
   \cmidrule{12-14}
%%%  No 9              
   9 & 2024-09-11 & 13814 & M1.8 & 400 & 12:27\,/\,12:36\,/\,12:47 & 12:31\,/\,13:22 & $\sim 80$\%& (146, & 130) & 0.979 & \CaIR & 35 & 7 \\
   \cmidrule{12-14}  
%%%  No 10                
   \multirow{2}{*}{10\tablefootmark{\ddag}} & \multirow{2}{*}{2014-09-06} & \multirow{2}{*}{12157} & \multirow{2}{*}{M1.1} & \multirow{2}{*}{340} & \multirow{2}{*}{16:50\,/\,17:09\,/\,17:22} & \multirow{2}{*}{15:27\,/\,17:27} & \multirow{2}{*}{$\sim100$\%} & \multirow{2}{*}{$(-732,$} & \multirow{2}{*}{$-302)$} & \multirow{2}{*}{0.560} & \CaIR & 11 & 6 \\
   & & & & & & & & & & & \Halpha & 11 & 6 \\%[1.5mm] 
   \cmidrule{12-14}   
%%%  No 11                      
   \multirow{2}{*}{11} & \multirow{2}{*}{2016-05-14} & \multirow{2}{*}{12543} & \multirow{2}{*}{C8.4} & \multirow{2}{*}{144} & \multirow{2}{*}{11:28\,/\,11:34\,/\,11:37} & \multirow{2}{*}{11:38\,/\,12:50} & \multirow{2}{*}{$\sim100$\%} & \multirow{2}{*}{(877,} & \multirow{2}{*}{$-66)$} & \multirow{2}{*}{0.384} & \CaIR & 12 & 4 \\
    & & & & & & & & & & & \Halpha & 12 & 4 \\%[1.5mm] 
   \cmidrule{12-14}
%%%  No 12
   \multirow{2}{*}{12\tablefootmark{\ddag}} & \multirow{2}{*}{2015-06-24} & \multirow{2}{*}{12371} & \multirow{2}{*}{C5.6} & \multirow{2}{*}{--} & \multirow{2}{*}{15:12\,/\,15:29\,/\,15:40} & \multirow{2}{*}{14:49\,/\,16:45} & \multirow{2}{*}{$\sim10$\%} & \multirow{2}{*}{(496,} & \multirow{2}{*}{175)} & \multirow{2}{*}{0.841} & \CaIR & 13 & 7 \\ 
   & & & & & & & & & & & \Halpha & 13 & 7  \\%[1.5mm] 
   \cmidrule{12-14}   
%%%  No 13                   
   \multirow{2}{*}{13\tablefootmark{\ddag}} & \multirow{2}{*}{2011-08-06} & \multirow{2}{*}{11267} & \multirow{2}{*}{C4.1} & \multirow{2}{*}{--} & \multirow{2}{*}{08:37\,/\,08:47\,/\,08:51} & \multirow{2}{*}{09:00\,/\,09:37} & \multirow{2}{*}{$\sim50$\%} & \multirow{2}{*}{$(-359,$} & \multirow{2}{*}{$-359)$} & \multirow{2}{*}{0.848} & \Fe{i 6302} & 28 & 3 \\ 
   & & & & & & & & & & & \Fe{i 5576} & 28 & 3 \\%[1.5mm] 
   \cmidrule{12-14}
%%%  No 14          
   14\tablefootmark{\dag} & 2015-05-05 & 12335  & C3.6 & -- & 11:55\,/\,11:58\,/\,12:00 & 11:57\,/\,12:01 & $\sim95$\% & $(-300,$ & $-190)$ & 0.929 & \HeD & 14 & 9 \\
   \cmidrule{12-14}
%%%  No 15                   
   \multirow{2}{*}{15} & \multirow{2}{*}{2022-06-26} & \multirow{2}{*}{13040} & C1.9 & \multirow{2}{*}{--} & 07:44\,/\,07:50\,/\,07:55 & 07:37\,/\,08:44 & \multirow{2}{*}{$\sim 90$\%} & \multirow{2}{*}{$(-135,$} & \multirow{2}{*}{$-235)$} & \multirow{2}{*}{0.959} & \CaIR & 40 & 3 \\ 
   & & & +C2.4 & & 08:00\,/\,08:12\,/\,08:20 & 07:37\,/\,08:57 & & & & & \Halpha & 40 & 3 \\%[1.5mm] 
   \cmidrule{12-14}
%%%  No 16     
   & & & & & & & & & & & \CaIR & 53 & 6 \\
   & & & & & & & & & & & \Halpha & 53 & 6 \\
   16\tablefootmark{\dag} & 2023-06-18 & 13336 & C2.2 & -- & 09:31\,/\,09:37\,/\,10:09 & 09:37\,/\,09:52 & $\sim100$\% & $(-407,$ & $-390)$ & 0.808 &  \HeD   & 53 & 6 \\
   & & & & & & & & & & & \Hbeta & 25 & 0 \\
   & & & & & & & & & & & \CaH & 25 & 0 \\%[1.5mm]        
   \cmidrule{12-14}
%%%  No 17
   & & & & & & & & & & & \CaIR  & 20 & 9 \\
   17 & 2019-05-06 & 12740 & C2.0 & 128 & 08:41\,/\,08:47\,/\,08:51 & 08:34\,/\,09:33 & $\sim100$\% & $(-717,$ & $186)$ & 0.638 & \Fe{i 6173} & 20 & 9 \\          
   & & & & & & & & & & & \CaK & 7 & 8 \\%[1.5mm] 
   \cmidrule{12-14}
%%%  No 18                        
   18 & 2013-06-30 & 11778 & C1.5 & -- & 09:11\,/\,09:18\,/\,09:27 & 09:16\,/\,10:17 & 100\% & (323, & $-288)$ & 0.892 & \Halpha & 7 & 3 \\
   \cmidrule{12-14}
%%%  No 19
   & & & & & & 09:49\,/\,10:16 & & & & & \CaIR & 19 & 2 \\
   19\tablefootmark{\dag} & 2021-05-26 & 12826 & C1.2 & -- & 09:46\,/\,09:55\,/\,10:00 & 09:49\,/\,10:16 & $\sim100$\% & (700, & 350) & 0.564 & \Fe{i 6173} & 19 & 2 \\
   & & & & & & 09:55\,/\,10:16 & & & & & \CaK & 7 & 3 \\%[1.5mm]
   \bottomrule
   \end{tabular}
   \tablefoot{
   \tablefoottext{\dag}{The data are archived in \href{https://dubshen.astro.su.se/sst_archive/search}{the Stockholm SST Archive.}}
   \tablefoottext{\ddag}{The data are archived in \href{https://star.pst.qub.ac.uk/wiki/public/solarflares/}{the QUB Solar flare database.}}
   \tablefoottext{a}{YYYY-MM-DD.}
   \tablefoottext{b}{From \href{https://www.lmsal.com/solarsoft/latest_events_archive.html}{SolarSoft Latest Events Archive.}} 
   \tablefoottext{c}{Linear Fit Speed from the \href{https://cdaw.gsfc.nasa.gov/CME_list/}{SoHO\,/\,LASCO CME Catalog} \citep{Gopalswamy_etal2009}.}    
   \tablefoottext{d}{Spatial overlap primarily estimated between the CRISP FOV and the total flare area in AIA 1700.}
   \tablefoottext{e}{The coordinates represent the average estimated center of the flare within the CRISP and/or CHROMIS FOV, determined by excluding highly variable regions and focusing on consistently covered areas.}}      
\end{table*}

\end{appendix}

\end{document}